\documentclass[onecolumn,9pt]{article}
\usepackage{extsizes}
\usepackage[super,sort&compress,comma]{natbib} 
\usepackage[version=3]{mhchem}
\usepackage[left=1.5cm, right=1.5cm, top=1.785cm, bottom=2.0cm]{geometry}
\usepackage{siunitx}
\usepackage{balance}
\usepackage{mathptmx}
\usepackage{sectsty}
\usepackage{graphicx} 
\usepackage{lastpage}
\usepackage[format=plain,justification=justified,singlelinecheck=false,font={stretch=1.125,small,sf},labelfont=bf,labelsep=space]{caption}
\usepackage{float}
\usepackage{fancyhdr}
\usepackage{fnpos}
\usepackage[english]{babel}
\addto{\captionsenglish}{%
  \renewcommand{\refname}{Notes and references}
}
\usepackage{array}
\usepackage{droidsans}
\usepackage{charter}
\usepackage[T1]{fontenc}
\usepackage[usenames,dvipsnames]{xcolor}
\usepackage{setspace}
\usepackage[compact]{titlesec}
\usepackage{hyperref}
\usepackage[version=3]{mhchem} 
\usepackage{booktabs}
\usepackage{longtable}
\usepackage{orcidlink}
\usepackage{rotating}

\usepackage{epstopdf}

\newcommand{\kcalmol}{\ensuremath{\text{kcal}\cdot\text{mol}^{-1}}}

\definecolor{cream}{RGB}{222,217,201}

\begin{document}

\pagestyle{fancy}
\thispagestyle{plain}
\fancypagestyle{plain}{
\renewcommand{\headrulewidth}{0pt}
}

\makeFNbottom
\makeatletter
\renewcommand\LARGE{\@setfontsize\LARGE{15pt}{17}}
\renewcommand\Large{\@setfontsize\Large{12pt}{14}}
\renewcommand\large{\@setfontsize\large{10pt}{12}}
\renewcommand\footnotesize{\@setfontsize\footnotesize{7pt}{10}}
\makeatother

\renewcommand{\thefootnote}{\fnsymbol{footnote}}
\renewcommand\footnoterule{\vspace*{1pt}%
\color{cream}\hrule width 3.5in height 0.4pt \color{black}\vspace*{5pt}} 
\setcounter{secnumdepth}{5}

\makeatletter 
\renewcommand\@biblabel[1]{#1}            
\renewcommand\@makefntext[1]%
{\noindent\makebox[0pt][r]{\@thefnmark\,}#1}
\makeatother 
\renewcommand{\figurename}{\small{Fig.}~}
\sectionfont{\sffamily\Large}
\subsectionfont{\normalsize}
\subsubsectionfont{\bf}
\setstretch{1.125} 
\setlength{\skip\footins}{0.8cm}
\setlength{\footnotesep}{0.25cm}
\setlength{\jot}{10pt}
\titlespacing*{\section}{0pt}{4pt}{4pt}
\titlespacing*{\subsection}{0pt}{15pt}{1pt}

\fancyfoot{}
\fancyfoot[LO,RE]{\vspace{-7.1pt}\includegraphics[height=9pt]{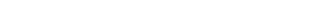}}
\fancyfoot[CO]{\vspace{-7.1pt}\hspace{13.2cm}\includegraphics{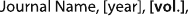}}
\fancyfoot[CE]{\vspace{-7.2pt}\hspace{-14.2cm}\includegraphics{head_foot/RF}}
\fancyfoot[RO]{\footnotesize{\sffamily{1--\pageref{LastPage} ~\textbar  \hspace{2pt}\thepage}}}
\fancyfoot[LE]{\footnotesize{\sffamily{\thepage~\textbar\hspace{3.45cm} 1--\pageref{LastPage}}}}
\fancyhead{}
\renewcommand{\headrulewidth}{0pt} 
\renewcommand{\footrulewidth}{0pt}
\setlength{\arrayrulewidth}{1pt}
\setlength{\columnsep}{6.5mm}
\setlength\bibsep{1pt}

\makeatletter 
\newlength{\figrulesep} 
\setlength{\figrulesep}{0.5\textfloatsep} 

\newcommand{\topfigrule}{\vspace*{-1pt}%
\noindent{\color{cream}\rule[-\figrulesep]{\columnwidth}{1.5pt}} }

\newcommand{\botfigrule}{\vspace*{-2pt}%
\noindent{\color{cream}\rule[\figrulesep]{\columnwidth}{1.5pt}} }

\newcommand{\dblfigrule}{\vspace*{-1pt}%
\noindent{\color{cream}\rule[-\figrulesep]{\textwidth}{1.5pt}} }

\makeatother

\twocolumn[
  \begin{@twocolumnfalse}
{\includegraphics[height=30pt]{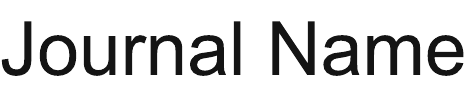}\hfill\raisebox{0pt}[0pt][0pt]{\includegraphics[height=55pt]{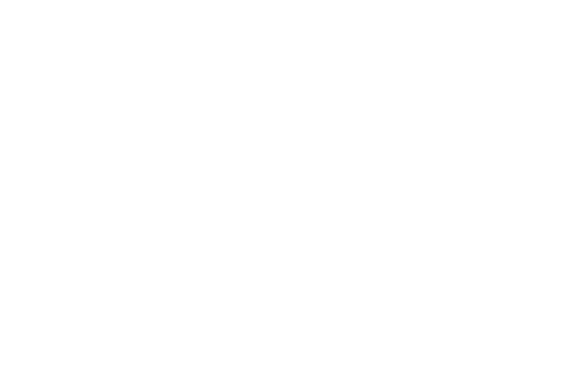}}\\[1ex]
\includegraphics[width=18.5cm]{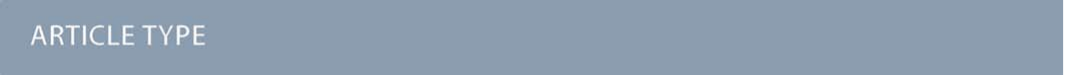}}\par
\vspace{1em}
\sffamily
\begin{tabular}{m{4.5cm} p{13.5cm} }

\includegraphics{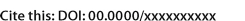} & \noindent\LARGE{\textbf{Reliable and Efficient Automated Transition-State Searches with Machine-Learned Interatomic Potentials}} \\
\vspace{0.3cm} & \vspace{0.3cm} \\

 & \noindent\large{Jonah Marks\orcidlink{0009-0006-7600-1095},\textit{$^{a}$} Jonathon Vandezande\orcidlink{0000-0001-7969-5498},\textit{$^{b}$} and Joseph Gomes\orcidlink{0000-0002-0755-0641}\textit{$^{a,*}$}} \\

\includegraphics{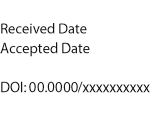} & \noindent\normalsize{Transition-state searches are central to understanding reaction mechanisms, but the high computational cost of density-functional theory (DFT) limits their application in high-throughput catalyst and materials discovery. Machine-learned interatomic potentials (MLIPs) offer near-DFT accuracy at orders-of-magnitude lower cost, yet their reliability for transition-state searches remains underexplored. Here, we systematically benchmark hybrid transition-state-search workflows combining six freely available potentials (MACE-OMol25, UMA-Small, UMA-Medium, eSEN-S, AIMNet2, and GFN2-xTB) with two reaction-path-finding algorithms (the freezing-string method and climbing-image nudged elastic band) across 58 diverse reactions spanning small organics, polymerization chemistry, and transition-metal catalysis. We find that models trained on the Open Molecules 2025 dataset exhibit markedly superior performance, with MACE-OMol25 achieving a 96.6\% success rate while requiring fewer than four DFT-gradient evaluations per reaction on organic systems---a 94–96\% reduction compared to conventional DFT-based searches. Low-level refinement on the MLIP surface before high-level DFT optimization reduces computational cost three-fold with minimal loss in reliability. For transition-metal systems, UMA-Medium demonstrates promising transferability to in-distribution transition metal complex reactions and out-of-distribution organometallic C–H activation. These results establish MLIP-accelerated workflows as practical tools for automated reaction discovery, enabling near–DFT accuracy at a fraction of traditional expense.} \\

\end{tabular}

 \end{@twocolumnfalse} \vspace{0.6cm}

  ]

\renewcommand*\rmdefault{bch}\normalfont\upshape
\rmfamily
\section*{}
\vspace{-1cm}


\footnotetext{\textit{$^{a}$~Department of Chemical and Biochemical Engineering, University of Iowa, Iowa City, IA, United States}}
\footnotetext{\textit{$^{b}$~Rowan Scientific, Boston, MA, United States}}
\footnotetext{\textit{$^{*}$}~Email: joe-gomes@uiowa.edu}



\setlength{\emergencystretch}{3em}
\section{Introduction}
Determining reaction mechanisms is central to the design of new materials, catalysts, and synthetic routes. Experimental elucidation of mechanisms and reaction parameters is often challenging and resource intensive due to the rarity of barrier-crossing events and the fleeting nature of activated complexes such as transition states. Theoretical approaches therefore play an essential complementary role, providing mechanistic insight and, once validated, may enable high-throughput screening of reactive systems. Many theoretical methods center on identifying the transition state, whose relative energy and geometry determine the rate and selectivity of a reaction. The transition state is the highest energy point along the minimum-energy path (MEP) connecting reactant and product states, corresponding to a first-order saddle point on the Born–Oppenheimer potential-energy surface (PES). The energy of the transition state relative to the reactant and product states defines the activation barrier and thereby the reaction kinetics. Accurately capturing these energetics requires PESs from electronic-structure methods such as density-functional theory (DFT), which can accurately describe complex chemical phenomena including bond formation and dissociation. However, despite favorable scaling relative to post-Hartree–Fock methods, DFT-based searches quickly become prohibitively expensive in large systems or when multiple reactions are under investigation due to the non-linear scaling of DFT and large number of in-series PES evaluations required by TS-search algorithms. Consequently, there is a growing need for workflows that can generate high-quality TS guesses with minimal computational cost.

To address this issue, many algorithms have been developed to automate TS searches, including machine-learning algorithms that attempt to directly predict reaction parameters or TSs\cite{grambow2020deep,Makos2021,Heinen2021,Spiekermann2022,van20243dreact,Duan2025,Chang2025} and methods based on a chain-of-states reaction-path representation that explore the PES. The latter group includes two functional classes: methods that optimize the MEP (such as the nudged elastic band and growing-string method), which provide approximate transition-state geometries, and local refinement techniques that take such geometries as starting points and converge them to true first-order saddle points, often using Hessian information. A widely used and effective strategy is to combine these approaches—first by employing a reaction-path locating method to generate a candidate transition state, followed by local refinement with a local surface-walking algorithm. \cite{Peters2004,behn2011efficient,MallikarjunSharada2012,Zimmerman2013,jafari2017reliable} A hybrid two-stage workflow can be used wherein the reaction-path-locating algorithm is performed at a lower level of theory, and the local refinement at a higher level of theory.

Among the various reaction-path-locating methods,\cite{mills1994quantum,henkelman2000improved,Henkelman2000climbing,Peters2004,sheppard2008optimization,behn2011efficient,Zimmerman2013} the nudged elastic band (NEB)\cite{mills1994quantum,henkelman2000improved,Henkelman2000climbing} and freezing-string method (FSM)\cite{behn2011efficient,MallikarjunSharada2012,Marks2024incorporation} are two of the most widely used approaches. The NEB method constructs a discretized reaction path between reactant and product via interpolation. The path is then converged to an approximate MEP through optimization of the spring-coupled images in the discretized path. The spring coupling ensures equidistant distribution of the images in the path. In the climbing-image nudged elastic band (CI-NEB) method, the highest-energy image is driven uphill along the reaction coordinate to converge to the saddle point, improving the accuracy of the transition state. In contrast to the NEB method, the FSM incrementally extends strings from the reactant and product states until two branches meet, offering improved convergence for complex pathways that deviate significantly from interpolated paths. FSM "freezes" earlier images in the path as new images are added, reducing the cost of the search at the expense of not finding the minimum energy pathway. Both techniques have been shown to perform well with \textit{ab-initio} PESs, but the substantial computational cost of repeated electronic-structure evaluations limits their practicality for large systems or high-throughput applications. Reducing the number of high-level evaluations required during transition-state searches is therefore key to improving the scalability of these methods.

Machine-learned interatomic potentials (MLIPs) have emerged as a compelling alternative to DFT, delivering near-DFT accuracy at significantly reduced computational cost. These models use neural networks or other machine-learning architectures to approximate the PES by learning from large datasets of quantum-chemical energies and forces, enabling rapid energy and gradient evaluations once trained. Recent model architectures such as AIMNet2,\cite{Anstine2025} MACE-OMol25,\cite{batatia2022mace} eSEN-S,\cite{Fu2025} and UMA\cite{wood2025family} offer orders-of-magnitude speedup relative to DFT while maintaining near-DFT accuracy across respective validation sets. With new quantum-chemical datasets like the Open Molecules 2025 (OMol25) dataset, pre-trained forms of MACE, eSEN-S, and UMA architectures are able to reliably simulate charged and open-shell systems with chemically diverse atomic species, enabling simulation of many classes of industrially-relevant molecules. \cite{Levine2025} While these MLIPs have been trained and tested extensively for thermochemical and structural properties, their performance in transition-state searches, where accurate description of the PES near the activated regions is critical, remains less explored. Prior applications of MLIPs in TS searches, often employing custom-trained models with narrow chemical domains, have shown promise but were typically limited to small benchmarks or specialized implementations of the reaction-path locating algorithms. \cite{Schreiner2022,Wander2024,Zhao2025,Marks2025}

Additionally, many machine-learning methods that directly predict transition states from reactant and product structures have been proposed. \cite{grambow2020deep,Duan2025} These methods are extremely promising, as they offer orders-of-magnitude cost reductions relative to the hybrid workflows presented here and often achieve high accuracy for reactions within their training distributions. However, their generalization to out-of-distribution reaction types remains limited. Development of workflows that integrate conventional search algorithms with MLIPs in a standardized framework enables researchers to rapidly incorporate emerging state-of-the-art models or train models for specific phenomena of interest, providing broader chemical coverage at the cost of higher per-reaction expense.

In this work, we develop and evaluate a hybrid machine-learned-DFT workflow for automated transition-state searches. MLIPs or semi-empirical DFT are first used to generate TS guess geometries using either the FSM or the CI-NEB method. Each potential–algorithm pair produces a candidate geometry that can optionally be refined on the same surface, hereafter referred to as low-level refinement. These native and low-level-refined guesses are then used as initial structures for high-level (DFT) refinement. This hybrid strategy allows the inexpensive low-level potential to bear the majority of the search cost, while the high-level refinement provides the accuracy and validation necessary for reliable thermochemical and kinetic predictions. \cite{Iron2019,bursch2022best}

We then systematically benchmark all combinations of the FSM and CI-NEB with six freely available potentials-MACE-OMol25, UMA-Small, UMA-Medium, eSEN-Small, AIMNet2 and GFN2-xTB- yielding 24 potential–algorithm workflows per reaction. Performance is assessed across two established reaction sets, followed by application of best-performing combinations to a new benchmark of polymerization reactions and four organometallic test cases. Two primary metrics are considered: the success rate of the high-level refinement to the reference saddle point as validated by frequency analysis, and the number of DFT gradient evaluations required to do so. Collectively these results reveal clear trends in which algorithm-potential combinations most effectively reproduce DFT-level transition states while minimizing high-level computational cost. Models trained on the OMol25 dataset exhibit markedly higher reliability and lower search costs than those trained on other datasets. Among these, MACE-OMol25 consistently achieves 96\% success rates across benchmark sets with an average of fewer than five DFT evaluations per reaction on small organic systems. The UMA-Medium model displays comparable efficiency, with slightly lower reliability on small molecules but improved performance for larger, more complex systems, including organometallic and transition-metal catalysis.

\section{Methods}
\subsection{Transition-State Searches}
\begin{figure}[h!]
    \centering
    \includegraphics[width=0.5\textwidth,trim=30cm 0cm 15cm 6cm, clip]{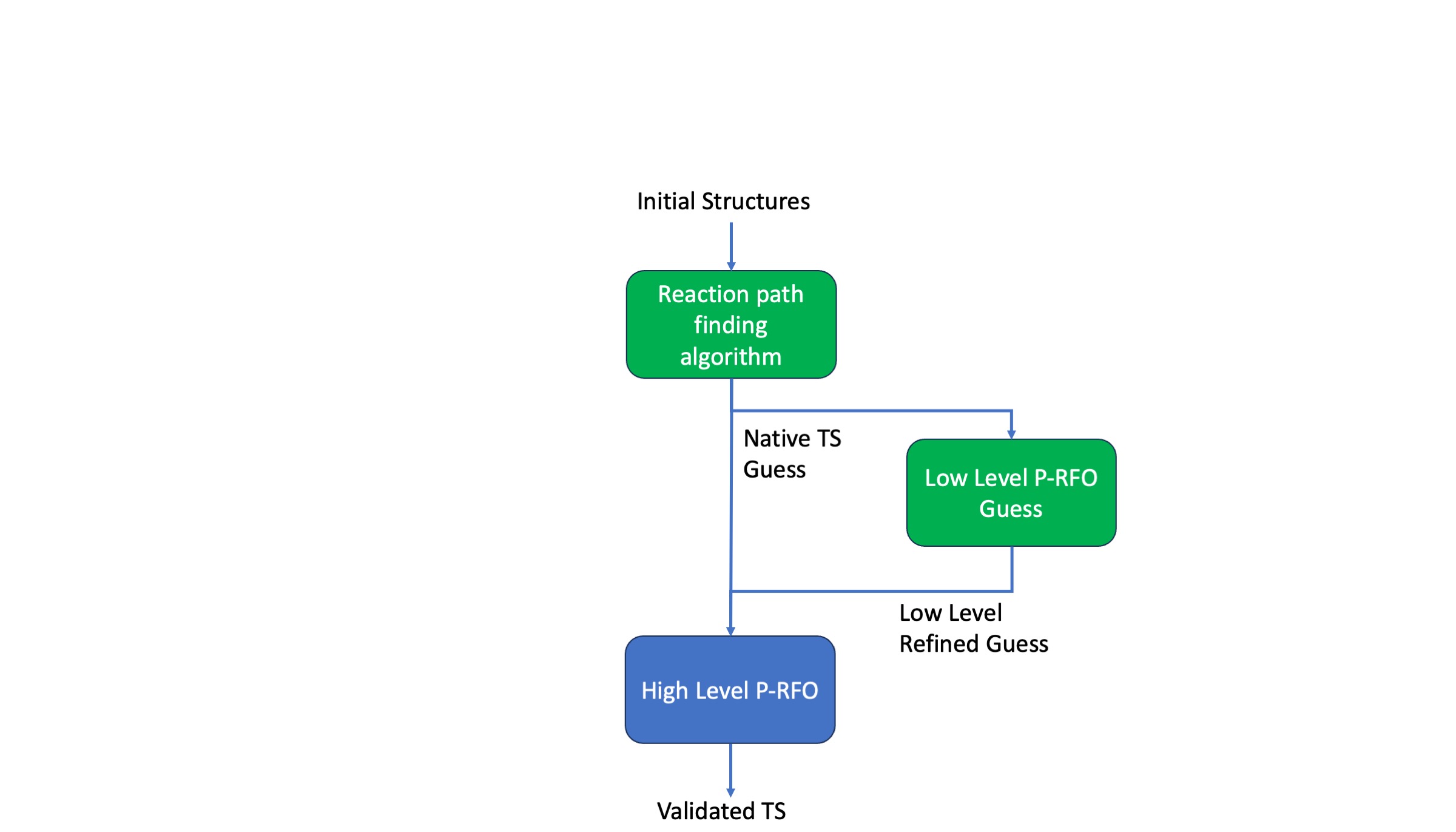}
    \caption{Transition-state-search workflows, green boxes denote steps performed with low-level potentials, while the blue box denotes use of the $\omega$B97X-V/def2-TZVP level of theory. }
    \label{fig:TS_workflows_sella}
\end{figure}

Transition-state-search workflows involve multiple stages, progressing from initial molecular structures to fully validated transition states. The key steps used in this study are outlined in Figure \ref{fig:TS_workflows_sella}. 

The reactant and product geometries provided by benchmark datasets and prior literature case studies were used as starting points for one of two reaction-path-finding algorithms: the ML-FSM Python package or the CI-NEB implementation in ASE\cite{larsen2017atomic}. The details and parameters of ML-FSM are described in section \ref{sec:methods_ML-FSM}; the CI-NEB implementation and parameters are described in Appendix~\ref{sec:methods_CI-NEB}. Both ML-FSM and CI-NEB are chain-of-states methods that produce a discrete string of optimized intermediate geometries connecting the reactant and product. The highest energy intermediate geometry, as computed on the low-level potential, is designated as the native transition-state guess.

Two workflow variants were evaluated. In the native workflow, this TS guess was passed directly to high-level refinement and validation using partitioned rational-function optimization in Q-Chem at the $\omega$B97X-V/def2-TZVP level of theory. In the low-level-refined workflow, the same TS guess was first refined on the low-level potential (MLIP or semi-empirical DFT) prior to the high-level refinement and validation (\textit{vide infra} for comparative analysis of both workflows). Low-level refinement was carried out using a restricted-step P-RFO optimization in redundant internal coordinates, as implemented in the Sella Python package.\cite{Hermes2019,Hermes2022} This additional refinement stage was intended to improve the proximity of the native guess to a first-order saddle point while retaining the efficiency of the low-level potential. 

\subsection{Non-local Path Finding Algorithms}
\subsubsection{ML-FSM}
\label{sec:methods_ML-FSM}
The Freezing-String Method (FSM) provides transition-state guesses at significantly reduced computational cost by employing a "freeze-and-grow" strategy that eliminates the need for global chain optimization.\cite{behn2011efficient} Unlike the GSM and NEB method, which iteratively optimize the entire chain of states, the FSM permanently "freezes" each node after its initial optimization, dramatically reducing the total number of gradient evaluations required at the cost of locating the MEP.

The FSM algorithm proceeds through iterative growth cycles consisting of two primary steps. First, interpolation is performed between the current innermost (frontier) nodes to generate a dense chain of intermediate states from which an intermediate geometry $\mathbf{R}^\text{interp}$ at fixed arc-length distance $s$ along the approximate reaction coordinate is selected. Second, this interpolated structure undergoes constrained optimization using a local quadratic approximation of the PES:

\begin{equation}
E(\mathbf{R}) = E(\mathbf{R}^{\text{interp}}) + (\mathbf{R} - \mathbf{R}^{\text{interp}})^T \mathbf{g}^{\perp} + \frac{1}{2}(\mathbf{R} - \mathbf{R}^{\text{interp}})^T \mathbf{H} (\mathbf{R} - \mathbf{R}^{\text{interp}})
\label{eq:fsm_quadratic}
\end{equation}

where $\mathbf{g}^\perp$ denotes the component of the gradient perpendicular to the local tangent direction and $\mathbf{H}$ is an approximate Hessian matrix. The perpendicular gradient is computed through the projection:

\begin{equation}
\mathbf{g}^{\perp} = (\mathbf{I} - \hat{\mathbf{t}}\hat{\mathbf{t}}^T)\mathbf{g}
\label{eq:fsm_perp_gradient}
\end{equation}

where $\hat{\mathbf{t}}$ is the normalized tangent vector at the current node and $\mathbf{I}$ is the identity matrix. 

The optimization is performed using the L-BFGS-B algorithm with backtracking line searches and user-specified limits on both the maximum number of line search steps and the maximum number of optimization steps that can be taken.\cite{Marks2024incorporation} Once optimized, each node is permanently frozen and excluded from further refinement, ensuring linear scaling with respect to the number of nodes added. 

Growth cycles continue until string unification is achieved. The unification criterion examines the distance $d$ between frontier nodes: if $2s > d > s$, a single node is added and the calculation terminates; if $d < s$ due to the optimization of frontier nodes, termination occurs without additional node placement. Upon completion, the highest energy image serves as the transition-state guess for subsequent refinement using local surface-walking methods. 

The key efficiency advantage of the FSM lies in the truncation of node optimization and absence of global chain optimization. While this approach sacrifices detailed minimum energy pathway information, it enables rapid identification transition-state region with computational costs typically 5–10 times lower than conventional GSM and NEB methods. \cite{behn2011efficient,Marks2024incorporation} However, unlike CI-NEB or GSM methods, the highest energy node cannot be considered the exact transition state. 

In this work, we use the ML-FSM Python package. FSM calculations used linear interpolation in redundant internal coordinates combined with Cartesian-space optimization using the L-BFGS-B implementation in SciPy.\cite{liu1989limited,byrd1995limited,virtanen2020scipy,Marks2024incorporation} Each calculation employed 18 nodes, a maximum of three line search steps, and two optimization steps per node. 

\subsection{Potential-Energy Surfaces}
\label{sec:MLIPS}
Computationally inexpensive PESs that maintain accuracy along reaction coordinates are essential for enabling high-throughput transition-state searches. Five freely available machine-learned interatomic potentials and one semi-empirical DFT (SE-DFT) models were used in this work for both reaction path finding and low-level refinement of the TS guesses. Each model is summarized below; full details can be found in the respective cited references. 

\begin{itemize}
    \item \textbf{GFN2-xTB}:\cite{bannwarth2019gfn2} An extended-tight-binding SE-DFT model with self-consistent multipole electrostatics and D4 dispersion. GFN2-xTB is parameterized up to Z=86.
    \item \textbf{AIMNet2}:\cite{Anstine2025} A second-generation atoms-in-molecules neural-network potential incorporating iterative charge equilibration and explicit long-range electrostatics. Trained on 20~$\omega$B97M-D3/def2-TZVPP structures distilled from a 120M B97-3c conformer master set of molecules with elements (C,H,N,O,S,F,Cl,Br,I,P,Si,B,As,Se).
    \item \textbf{oMOL25's eSEN Conserving small}:\cite{Levine2025} (hereafter eSEN-S): an equivariant graph transformer with spherical convolutions instead of attention trained on the OMol25 dataset.
    \item \textbf{UMA Small 1.1}:\cite{wood2025family} (hereafter UMA-S): mixture-of-experts variants of eSEN-S with 150M total parameters ($\approx 6$M active) trained on approximately 500M structures.\cite{Chanussot2021,Barroso-Luque2024,Sriram2024,Levine2025}
    \item \textbf{UMA Medium 1.1}:\cite{wood2025family} (hereafter UMA-M): mixture-of-experts variants of eSEN-S with 1.4B parameters ($\approx 150$M active) trained on approximately 500M structures.\cite{Chanussot2021,Barroso-Luque2024,Sriram2024,Levine2025}
    \item \textbf{MACE-OMol25}:\cite{batatia2022mace} An $E(3)$-equivariant message-passing neural network with explicit higher-body terms (up to 4-body) and bilinear radial mixing. The OMol25 variant introduces charges and spin awareness, trained on $\sim$100M DFT data points ($\omega$B97M-V/def2-TZVPD) spanning diverse charged and open-shell molecules. 
\end{itemize}

\subsection{Computational Details}
All DFT calculations were performed with Q-Chem 6.0,\cite{Epifanovsky2021} using the range-separated-hybrid density functional $\omega$B97X-V \cite{Mardirossian2014} with the triple-$\zeta$ def2-TZVP\cite{Weigend2005} basis set. The SG-2 quadrature grid was employed for all energy and gradient evaluations. \cite{dasgupta2017standard} In several cases across benchmark sets, TS searches converged to the correct saddle point with a secondary imaginary frequency, often below $20\text{i}$. To eliminate these secondary imaginary frequencies the optimization was continued with the SG-3 quadrature grid, tighter SCF convergence criteria, and tighter convergence thresholds.

DFT calculations were primarily used for two tasks: (1) harmonic-frequency calculation to verify the nature of saddle points, where a first-order saddle point must have exactly one imaginary frequency, and (2) the high-level refinement of guess geometries to first-order saddle points using the eigenvector following partitioned rational-function optimization (P-RFO)\cite{Baker1986} method as implemented in Q-Chem. The P-RFO algorithm operates in delocalized internal coordinates by default. In instances where P-RFO converged to an off-target saddle point, the connected minima were confirmed by intrinsic reaction coordinate (IRC) calculations performed in Sella using the predictor–corrector geodesic algorithm.\cite{schmidt1985intrinsic,Hermes2022}

Refinement of native TS guesses on the low-level PES—either a MLIP or SE-DFT—was carried out using a restricted-step partitioned-rational-function optimization (RS-P-RFO) in redundant internal coordinates, as implemented in the Sella Python package.\cite{Hermes2022} The RS-P-RFO employs iterative Hessian diagonalization enabling efficient convergence to nearby first-order saddle points without full Hessian evaluations.\cite{Hermes2019} Further details of the optimization algorithm are provided in the original works \citet{hermes2021geometry,Hermes2022}. Convergence was achieved when the maximum Cartesian gradient component was below 0.05~eV$\cdot$\AA$^{-1}$ (approximately $3\times 10^{-3}~\text{Ha}\cdot\text{Bohr}^{-1}$). Transition-state geometries refined at the low-level were subsequently used as initial guesses for the high-level ($\omega$B97X-V/def2-TZVP) P-RFO refinement and validation.

High-level transition-state refinement optimizations performed in Q-Chem 6.0 are initialized with an analytical Hessian that is updated at each cycle using the Murtagh–Sargent–Powell formula.\cite{murtagh1970computational} The maximum number of self-consistent-field (SCF) iterations and geometry-optimization steps were limited to 250. SCF energies and gradients were converged to $10^{-6}$ Ha and $3\times10^{-4}~\text{Ha}\cdot\text{Bohr}^{-1}$, respectively. In cases where frequency analysis revealed low-magnitude ($<30 \text{ cm}^{-1}$) or spurious imaginary modes, tighter convergence criteria was employed to eliminate minor imaginary modes.

\section{Results}
\subsection{Organic Reaction Benchmarking}
To evaluate the reliability and efficiency of the MLIPs introduced in Section~\ref{sec:MLIPS} for the transition-state searches, we applied four workflow variants (FSM native, FSM low-level refined, CI-NEB native, and CI-NEB low-level refined) with all six potentials to two established benchmark sets of organic reactions. The CI-NEB results are discussed in detail in Appendix~\ref{sec:FSM_NEB_COMPARE}; here we focus on the FSM-based searches, which achieved higher success rates at lower computational cost. The Baker set, introduced by \citet{baker1996location}, consists of 24 gas-phase reactions of small organic molecules. The reactions include chemical phenomena such as insertion reactions, cationic and anionic mechanisms, open-shell systems, and rotation reactions, providing a rigorous test of algorithm robustness across diverse organic reaction types. The second set, introduced by \citet{MallikarjunSharada2012}, comprises nine elementary organic reactions encompassing diverse reaction phenomena including bond formation, dissociation, ring-opening, and isomerization. Despite its modest size, the set includes both prototypical and challenging cases–for example the Alanine Dipeptide, parent Diels–Alder, and Ireland–Claisen reactions, as well as a classically-hard test case, ethane dehydrogenation ($\text{CH}_3\text{CH}_3\rightarrow\text{CH}_2\text{CH}_2 +\text{H}_2$). Collectively, the Baker and Sharada benchmarks span a broad range of chemical transformations and charge-spin states while remaining tractable for high-level electronic-structure methods, making them ideal benchmark sets for evaluating transition-state-search methods.\cite{Heyden2005,kastner2008superlinearly,MallikarjunSharada2012,Marks2024incorporation,Marks2025,Hait2025}

\subsubsection{FSM}\label{sec:fsm_organic}

\begin{sidewaystable*}
    \centering
    \caption{Comparison of performance of GFN2-xTB, AIMNet2, eSEN-S, UMA-S, UMA-M, MACE-OMol25 for native and low-level-refined guess generation via the FSM on the Baker set. Performance is measured by successful convergence to the reference transition state, and the number of DFT gradient evaluations required. Italicized values denote failed runs, with superscripts denoting the failure mode.}
    \label{tab:fsm_baker}
    \resizebox{\textwidth}{!}{
      \begin{tabular}{l cc cc cc cc cc cc}
        \toprule
         & \multicolumn{2}{c}{GFN2-xTB} & \multicolumn{2}{c}{AIMNet2} & \multicolumn{2}{c}{eSEN-S} & \multicolumn{2}{c}{UMA-S} & \multicolumn{2}{c}{UMA-M} & \multicolumn{2}{c}{MACE-OMol25} \\
         &  {Native} & {Low-level} & {Native} & {Low-level} & {Native} & Low-level & {Native} & {Low-level} &  {Native} & {Low-level} &  Native & {Low-level} \\
         Reaction & &{refined}& &{refined}& &{refined}& &{refined}& &{refined}& &{refined}\\
        \midrule
        1. HCN $\rightarrow$ HNC & 3 & 4 & 3 & 4 & 4 & 2 & 3 & 3 & 4 & 2 & 4 & 2 \\
        2. HCCH $\rightarrow$ CCH$_2$ & 16 & 5 & 6 & 3 & 8 & 2 & 9 & 2 & 9 & 2 & 9 & 3 \\
        3. H$_2$CO $\rightarrow$ H$_2$ + CO & 10 & 5 & 21 & {\textit{36}$^b$} & 17 & 2 & 16 & 3 & 16 & 2 & 18 & 2 \\
        4. CH$_3$O $\rightarrow$ CH$_2$OH & 6 & 4 & 23 & 14 & 6 & 3 & 6 & 3 & 6 & 3 & 6 & 3 \\
        5. cyclopropyl ring opening & 13 & 5 & {\textit{47}$^a$} & {\textit{7}$^a$} & 13 & 3 & 12 & 3 & 9 & 3 & {\textit{23}$^a$} & 3 \\
        6. bicyclo[1.1.0]butane $\rightarrow$ $\textit{trans}$-butadiene & {\textit{147}$^a$} & 14 & 43 & 28 & 59 & 4 & {\textit{181}$^a$} & 4 & 43 & 3 & {\textit{92}$^a$} & 4 \\
        7. formyloxyethyl 1,2-migration & 14 & 8 & 18 & {\textit{33}$^b$} & 14 & 4 & 15 & 5 &11 & 4 & 16 & 3 \\
        8. parent Diels–Alder cycloaddition & 31 & 7 & 29 & 4 & 42 & 2 & 26 & 3 &19 & 3 & 23 & 3 \\
        9. s-tetrazine $\rightarrow$ 2HCN + N$_2$ & 14 & {\textit{10}$^b$} & 7 & 4 & 24 & 5 & 36 & 4 &24 & 3 & 9 & 3 \\
        10. $\textit{trans}$-butadiene $\rightarrow$ $\textit{cis}$-butadiene & 5 & 4 & 4 & 4 & 3 & 3 & 3 & 2 & 3 & 2 & 3 & 2 \\
        11. CH$_3$CH$_3$ $\rightarrow$ CH$_2$CH$_2$ + H$_2$ & 20 & 9 & 33 & 34 & 23 & {\textit{250}$^d$} & 20 & {\textit{156}$^a$} &17 & {\textit{64}$^a$} & 21 & 3 \\
        12. CH$_3$CH$_2$F $\rightarrow$ CH$_2$CH$_2$ + HF & 11 & 7 & 9 & 4 & 11 & 2 & 11 & 2 &11 & 2 & 11 & 2 \\
        13. acetaldehyde keto-enol tautomerism & 6 & 5 & 4 & 4 & 4 & 2 & 4 & 2 & 4 & 2 & 4 & 2 \\
        14. HCOCl $\rightarrow$ HCl + CO & 8 & 6 & 8 & 4 & 7 & 2 & 6 & 2 & 7 & 2 & 7 & 2 \\
        15. H$_2$O + PO$_3^-$ $\rightarrow$ H$_2$PO$_4^-$ & 23 & 6 & 27 &  {\textit{250}$^{d}$} & 23 & 3 & 27 & 2 & 21 & 2 & 23 & 2 \\
        16. CH$_2$CHCH$_2$CH$_2$CHO Claisen rearrangement & 30 & 7 & 34 & 6 & 27 & 22 & 24 & {\textit{250}$^{d}$} & 27 & {\textit{233}$^{a}$} & 25 & 3 \\
        17. SiH$_2$ + CH$_3$CH$_3$ $\rightarrow$ SiH$_3$CH$_2$CH$_3$ & 19 &22 & 7 & 8 & 6 & 4 & 7 & 5 & 7 & 4 & 6 & 4 \\
        18. HNCCS $\rightarrow$ HNC + CS & {\textit{144}$^a$} & {\textit{36}$^b$} & 8 & 4 & 21 & 3 & 12 & 3 &15 & 3 & 10 & 2 \\
        19. HCONH$_3^+$ $\rightarrow$ NH$_4^+$ + CO & {42$^c$} & 9 & {\textit{250}$^{d}$} & 21 & {14$^c$} &  {10$^c$} & {25$^c$} &  {9$^c$} & {17$^c$} & {9$^c$} & 21 & 7 \\
        20. acrolein rotational TS & 4 & 4 & 3 & 3 & 4 & 2 & 4 & 3 & 4 & 2 & 4 & 3 \\
        21. HCONHOH $\rightarrow$ HCOHNHO & 6 & 10 & 9 & 7 & 9 & 3 & 9 & 3 & 10 & 3 & 10 & 3 \\
        22. HNC + H$_2$ $\rightarrow$ H$_2$CNH & {\textit{45}$^{a}$} & 8 & {\textit{0}$^e$} & {\textit{17}$^a$} & {\textit{250}$^{d}$} & 2 & {\textit{250}$^{d}$} & {\textit{40}$^{a}$} & {\textit{41}$^a$} & 2 & {\textit{234}$^b$} & 3 \\
        23. H$_2$CNH $\rightarrow$ HCNH$_2$ & 9 & 7 & 7 & 10 & 6 & 2 & 5 & 2 & 6 & 3 & 5 & 3 \\
        24. HCNH$_2$ $\rightarrow$ HCN + H$_2$ & {\textit{38}$^a$} & {\textit{3}$^a$} & 37 & 29 & 40 & {\textit{18}$^f$} & 29 & {\textit{19}$^f$} & 28 & {\textit{18}$^f$} & 33 & {\textit{27}$^f$} \\
        \midrule
        Success Rate & 83.3\% & 87.5\% & 87.5\% & 79.2\% & 95.8\% & 91.7\% & 91.7\% & 83.3\% & 95.8\% & 87.5\% & 87.5\% & 95.8\% \\
        Mean Success Cost & 14.5& 7.4 & 16.2 & 10.3 & 16.7 & 4.0 & 14.0 & 3.3 & 13.8 & 2.9 & 12.8 & 2.9 \\
        \bottomrule
      \end{tabular}
    }
    \raggedright
    (a) Converges to an alternate first-order saddle point. (b) Converges to local-minimum structure. (c) Converges to correct TS with spurious imaginary frequency additional cost to eliminate frequency via tighter P-RFO convergence parameters was added. (d) Fails to converge P-RFO within 250 optimization cycles. (e) SCF convergence error. (f) Converges to structure with multiple strong ($>200 \text{ cm}^{-1}$) imaginary frequencies.\\
\end{sidewaystable*}

\begin{sidewaystable}
    \centering
    \caption{Comparison of performance of GFN2-xTB, AIMNet2, eSEN-S, UMA-S, UMA-M, MACE-OMol25 for native and low-level-refined guess generation via the FSM on the Sharada set. Performance is measured by successful convergence to the reference transition state, and the number of DFT gradient evaluations required. Italicized values denote failed runs, with superscripts denoting the failure mode.}
    \label{tab:FSM_sharada}
    \resizebox{\textwidth}{!}{\begin{tabular}{l cc cc cc cc cc cc}
        \toprule
         & \multicolumn{2}{c}{GFN2-xTB} & \multicolumn{2}{c}{AIMNet2} & \multicolumn{2}{c}{eSEN-S} & \multicolumn{2}{c}{UMA-S} & \multicolumn{2}{c}{UMA-M} & \multicolumn{2}{c}{MACE-OMol25} \\
         &  Native & Low-level &  Native & Low-level &  Native & Low-level &  Native & Low-level &  Native & Low-level &  Native & Low-level \\
         Reaction & &refined& &refined& &refined& &refined& &refined& &refined\\
        \midrule
        1. H$_2$CO $\rightarrow$ H$_2$ + CO & 10 & 5 & 21 & \textit{36}$^b$ & 17 & 2 & 16 & 3 &16 & 2 & 18 & 2 \\
        2. SiH$_2$ + H$_2$ $\rightarrow$ SiH$_4$ & 25 &  20 & 6 & 8 & 7 &  4 &  5 &  5 & 7 &  4 & 6 &  4 \\
        3. CH$_2$CHOH $\leftrightarrow$ CH$_3$CHO &  6 & 5 & 4 & 4 & 4 &  2 &  4 &  2 & 4 &  2 & 4 &  2 \\
        4. CH$_3$CH$_3$ $\rightarrow$ CH$_2$CH$_2$ + H$_2$ & 20 & 9 & 33 & 34 & 23 &  \textit{250}$^d$ & 20 &  \textit{156}$^a$ &17 & \textit{64}$^a$ & 21 & 3 \\
        5. bicyclo[1.1.0]butane $\rightarrow$ $\textit{trans}$-butadiene  &  \textit{147}$^a$ &14 & 43 & 28 & 59 & 4 &  \textit{181}$^a$ & 4 &43 & 3 & \textit{92}$^a$ & 4 \\
        6. parent Diels–Alder cycloaddition & 31 & 7 & 29 & 4 & 42 & 2 & 26 & 3 &19 & 3 & 23 & 3 \\
        7. cis,cis-2,4-hexadiene $\leftrightarrow$ 3,4-dimethylcyclobutene & 18 & 5 & 19 & 5 & 16 &  3 & 16 &  3 & 16 &  3 & 17 &  3 \\
        8. C$_5$ $\leftrightarrow$ C$_{7AX}$ & 153 &  23 & 100$^c$ & 25 &  108 & 12 & 65 & 12 &  199 & 12 &  117 & 16 \\
        9. silyl ketene acetal $\rightarrow$ silyl ester Ireland–Claisen rearrangement & \textit{250}$^d$ &  \textit{53}$^b$ &  107 &  19 &  49 & 14 & \textit{96}$^g$ & 15 &  75 &  \textit{250}$^d$ & 80 & 14 \\
        \midrule
        Success Rate & 77.8\% & 88.9\% & 100.0\% & 88.9\% & 100.0\% & 88.9\%& 77.8\% & 88.9\% & 100.0\% & 77.8\% & 88.9\% & 100.0\% \\
        Mean Success Cost & 37.6 & 11.0 & 40.2 & 15.9 & 36.1 & 5.4 & 21.7 & 5.9 & 44.0 & 4.1 & 35.8 & 5.7 \\
        \bottomrule
    \end{tabular}
    }
    \raggedright
    (a) Converges to an alternate first-order saddle point. (b) Converges to local-minimum structure. (c) Converges to correct TS with spurious imaginary frequency additional cost to eliminate frequency via tighter P-RFO convergence parameters was added. (e) SCF convergence error. (f) Converges to structure with multiple strong ($>200 \text{ cm}^{-1}$) imaginary frequencies. (g) P-RFO optimization step failure.\\
\end{sidewaystable}

Across the Baker benchmark, the FSM showed consistently high success rates and low computational cost in identifying reference saddle points. Averaged across all workflows and potentials, the FSM achieves a success rate of 88.9\%; successful searches required an average of 9.9 gradient evaluations. Native workflows, in which the FSM guess was submitted directly to the Q-Chem P-RFO, yielded the highest success rate at 90.3\% but required 14.7 gradient evaluations per successful search. The low-level refinement yielded a success rate of 87.5\% across all six models and the cost reduced by approximately 3-fold, with an average of 5.0 gradient evaluations per successful search—a 66\% reduction in cost. Under the assumption that the low-level potentials produce qualitatively correct Hessians, such a reduction in cost is expected. The modest decline in success rate, however, suggests that the low-level models do not yield uniformly accurate Hessians and in some cases introduce systematic failure. For example, in reaction 11 (CH$_3$CH$_3$ $\rightarrow$ CH$_2$CH$_2$ + H$_2$), searches using eSEN-S, UMA-S, and UMA-M all succeed with the native guess, but all fail after the low-level refinement. Likewise, for reaction 24 (HCNH$_2$ $\rightarrow$ HCN + H$_2$), eSEN-S, UMA-S, UMA-M, and MACE-OMol25 searches succeed natively yet fail in identical fashion following refinement. These two reactions represent characteristic failure modes and are analyzed in greater detail \emph{vide infra} in Section~\ref{sec:general_failure}.

Performance on the Sharada benchmark set mirrored that of the Baker set. Across the 9 reactions, 4 of which are shared with the Baker set, the FSM achieved an overall success rate of 89.8\% with an average cost of 22.4 gradient evaluations per successful search. The native and low-level-refined workflows achieved success rates of 90.7\% and 88.9\% respectively, with average costs of 36.4 and 8.0 gradient evaluations, respectively. Notably, the Sharada set includes larger systems like the alanine dipeptide isomerization (reaction 8) and the Ireland–Claisen rearrangement (reaction 9). The continued high performance on these 20–50+ atom systems demonstrates the FSM's scalability to reactions with many orthogonal degrees of freedom. The more pronounced cost reduction in this set underscores the efficiency of coupling the FSM with a local refinement algorithm prior to high-level refinement and validation. For larger systems, where the non-linear scaling of DFT renders gradient evaluations significantly more expensive, this reduction in the number of required evaluations becomes increasingly impactful. 


\subsubsection{Summary of MLIP performance on organic benchmarks}

When comparing machine-learned potentials, a strong and systematic performance advantage was observed for models trained on the OMol25 dataset. Evaluated over 29 unique reactions (24 from the Baker set and 5 from the Sharada set) using the FSM with low-level refinement, MACE-OMol25 achieved the highest success rate of 96.6\% with a mean cost of 3.8 gradient evaluations per successful search. eSEN-S followed closely (93.1\%, 4.5 gradients), while UMA-M achieved the lowest computational cost (3.3 gradients) at a moderate success rate of 86.2\%. UMA-S and GFN2-xTB showed comparable reliability (86-88\%) and AIMNet2 had the lowest success rate at 82.8\% with a mean cost of 10.7 gradient evaluations per successful search. 

The OMol25 trained models appear to have a systematic advantage. Aggregated across the full benchmark with both workflows, the four OMol25 models achieved a mean success rate of 91.8\% with an average cost of 11.7 gradient evaluations, compared to 84.5\% success and 15.5 gradient evaluations for non-OMol25 models. While training strategies, model architecture, and model hyperparameters influence the end model's accuracy, this difference likely arises from the richer representation of reactive configurations and transition-state geometries in the OMol25 dataset. Based on these results, the MACE-OMol25 and UMA-M combined with the FSM were selected for all subsequent evaluations of benchmark sets and case studies. 

\subsection{Closed-Shell Polymerization Reactions (Poly25)}

\begin{table*}[t]
\centering
\caption{Comparison of performance of UMA-M and MACE-OMol25 for native and low-level-refined guess generation via the FSM on the Poly25 benchmark. Performance is measured by successful convergence to the reference transition state and the number of DFT gradient evaluations required.}
\label{tab:long_results}

\begin{tabular}{l S S S S}
\toprule
Reaction  & \multicolumn{2}{c}{UMA-M} & \multicolumn{2}{c}{MACE-OMol25} \\
\cmidrule(lr){2-3} \cmidrule(lr){4-5}
 & {Native} & {Low-level Refined} & {Native} & {Low-level Refined} \\
\midrule

\textbf{Amide Formation} & & & & \\
\midrule
1. Methylamine + Benzoic acid & 205 & 6 & 46 & 5 \\
2. F$_3$-methylamine + F$_3$-acetic acid & 18 & 3 & 14 & 3 \\
3. Methylamine + F$_3$-acetic acid & 54 & \textit{250} & 55 & 4 \\
4. Methylamine + Acetic acid & 29 & 6 & 24 & 6 \\
5. F$_3$-methylamine + Acetic acid & 37 & 4 & 32 & 4 \\
6. Phenylamine + Acetic acid & 16 & 4 & 11 & 5 \\

\midrule
\textbf{Epoxy} & & & & \\
\midrule
7. Ethylene oxide + \textit{tert}-butoxide & 9 & 6 & 6 & 5 \\
8. Ethylene oxide + Methoxide & 8 & 5 & 6 & 4 \\
9. Fluoroethylene oxide + Methoxide & 27 & 5 & 37 & 5 \\
10. Propylene oxide + Methoxide & 7 & 5 & 7 & 8 \\

\midrule
\textbf{Ester Formation} & & & & \\
\midrule
11. F$_3$-methanol + F$_3$-acetic acid & 25 & 3 & 19 & 2 \\
12. Phenol + Acetic acid & 20 & 6 & 20 & 6 \\
13. Methanol + Acetic acid & 32 & 10 & 41 & 9 \\
14. Methanol + F$_3$-acetic acid & 30 & 4 & 11 & 5 \\
15. F$_3$-methanol + Acetic acid & 45 & 4 & 33 & 5 \\
16. Methanol + Benzoic acid & 46 & 9 & 35 & 8 \\

\midrule
\textbf{Isocyanate Containing} & & & & \\
\midrule
17. Isocyanate + Methanol & \textit{136} & \textit{250} & \textit{83} & \textit{250} \\
18. Isocyanate + Water & 8 & 2 & 7 & 3 \\
19. Methyl isocyanate + Water & 11 & 4 & 12 & 4 \\
20. Isocyanate + Phenol & 14 & 5 & 21 & 5 \\
21. Phenyl isocyanate + Water & 46 & 3 & 30 & 4 \\

\midrule
\textbf{Polyurea Formation} & & & & \\
\midrule
22. Isocyanate + Ammonia & 12 & 2 & 8 & 2 \\
23. Isocyanate + N-phenylaniline & 14 & 6 & 14 & 8 \\
24. Ethylene oxide + Phenoxide & 11 & 8 & 11 & 10 \\
25. Isocyanate + Aniline & 32 & 5 & 28 & 5 \\

\midrule
Success Rate & 96.0\% & 92.0\% & 96.0\% & 96.0\% \\
Mean Success Cost & 31.5 & 5.0 & 22.0 & 5.2 \\
\bottomrule

\end{tabular}
\end{table*}

The Poly25 dataset consists of 25 closed-shell gas phase reactions important in polymer chemistry. The reactions are further sub-divided into five classes: amide formation, epoxy polymerization, ester formation, isocyanate-containing reactions, and polyurea formation. The utility of this set of reactions is that it offers an assessment of the performance of current models and methods on an industrially relevant class of chemical reactions, that importantly has not been used in previous studies benchmarking reaction path and TS-finding methods. We apply the FSM-based native optimization and low-level refinement workflows using the MACE-OMol25 and UMA-M models to the problem of identifying the reference TS structures for each reaction contained in the Poly25 dataset.

Both MACE-OMol25 and UMA-M demonstrate excellent performance on the closed-shell-polymerization benchmark, achieving 96.0\% success rates–comparable or better than their respective performance on the combined Baker and Sharada sets, indicating that closed-shell polymerization reactions are an area of chemical space well represented in the OMol25 dataset. The mean computational cost for successful searches differs substantially between native workflows, with MACE-OMol25 averaging 22.0 gradient evaluations while UMA-M averages 31.5 gradient evaluations. The low-level refinement reduces both models to nearly identical costs of approximately 5 gradient evaluations. One notable outlier is reaction 1, methylamine and benzoic acid, where the UMA-M direct search requires 205 gradient evaluations compared to only 46 for MACE-OMol25. However, low-level refinement mitigates this disparity entirely, reducing both searches to 6 and 5 gradient evaluations respectively. 

For three of the four workflow combinations, UMA-M native, and both MACE-OMol25 native and low-level-refined searches, the sole failure occurs on reaction 19, the isocyanate and methanol reaction. Closer inspection of the FSM output reveals a discontinuous arc length along the internal coordinates interpolation path arising from failures in the back transformation from internal coordinates to Cartesian coordinates that occurs during the FSM growth phase. Manually re-running this reaction with linear synchronous transit (LST) interpolation,\cite{Halgren1977} a well-established alternative to internal coordinates interpolation for chain-of-states methods, circumvents this issue entirely. UMA-M direct and Sella-refined searches converge to the reference saddle point in 17 and 3 gradient evaluations respectively, while MACE-OMol25 requires 12 and 4 gradient evaluations. This failure mode can therefore be attributed to the ML-FSM implementation rather than deficiencies in the underlying potentials. While incorporation of fallback routines for common failure modes represents a good practices in high throughput workflow development, for the present analysis, this reaction is classified as a failure of the ML-FSM algorithm. The UMA-M low-level-refined workflow exhibits one additional failure on reaction 8, methylamine and F$_3$-acetic acid, reaching the 250 optimization step limit, whereas the MACE-OMol25 low-level-refined structure successfully converges in 4 steps.

\subsection{Organometallic Case Studies}
\begin{figure*}
\includegraphics[width=0.9\textwidth,trim=0cm 10cm 0cm 0cm, clip]{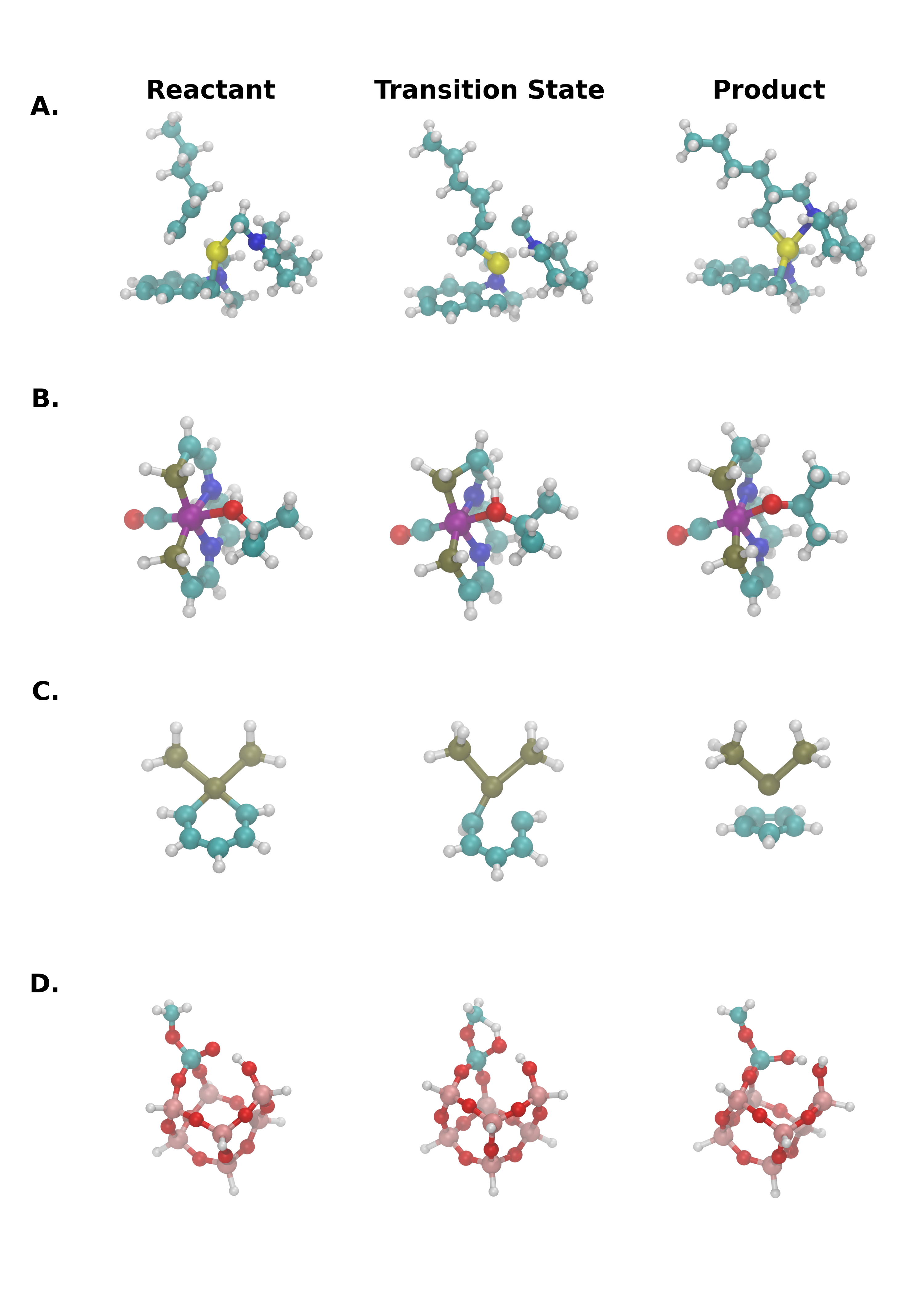}
\caption{Reactant, transition state, and product geometries of (A) Scandium-Catalyzed Olefin Insertion (B) Iron-Catalyzed Transfer Hydrogenation (C) Platinum-Mediated Metallabenzene Re-arrangement case studies.}
\label{fig:Case_studies}
\end{figure*}

Transition-metal-catalyzed reactions represent a critical test for MLIP-accelerated transition-state searches. These systems typically contain 30–200+ atoms, involve diverse metal-ligand coordination environments, and exhibit multiple competing reaction pathways, making conventional DFT-based mechanistic studies computationally demanding and well suited candidates for MLIP acceleration. To evaluate the UMA-M and MACE-OMol25 models on transition-metal-containing systems, we selected four case studies spanning different metals, coordination geometries, and mechanistic classes (Figure~\ref{fig:Case_studies}). 

The first three reactions are drawn from the Metal-Organic Barrier Heights (MOBH35) benchmark database, which provides DLPNO-CCSD(T)/CBS reference energies for 35 transition-metal-catalyzed transformations.\cite{Iron2019} The OMol25 training set includes reactive configurations generated from the MOBH35 structures via metal and ligand substitution and reaction path sampling with the Artificial-Force-Induced Reaction (AFIR) scheme.\cite{maeda2016artificial,Levine2025} The three selected MOBH35 reactions (scandium-catalyzed olefin insertion, iron-catalyzed transfer hydrogenation, and platinum-mediated metallabenzene rearrangement) therefore represent near-in-distribution test cases, assessing whether coverage of structurally related reactive configurations in the training data translates to reliable transition-state search performance. The three selected reactions span early and late transition metals, diverse coordination geometries, and distinct mechanistic classes. 

Following the MOBH35 test cases, we examine the rhodium(III)-catalyzed oxidative carbonylation of toluene, a well-characterized example of transition-metal-mediated C–H activation.\cite{Zakzeski2009} This system provides an out-of-distribution test case, as it is not derived from any of the reaction datasets used to generate OMol25 training configurations, probing MLIP transferability beyond the training data distribution. 

For all organometallic test cases, we applied the FSM native and low-level-refined workflows using the UMA-M and MACE-OMol25 models. Where experimental or high-level (DLPNO-\allowdisplaybreaks CCSD(T)/CBS) data was available we additionally evaluated the accuracy of MLIP predicted reaction and activation energies. 

\subsubsection{Scandium-Catalyzed Olefin Insertion}

The first case study examines the elementary olefin-insertion step in the scandium-catalyzed hydroaminoalkylation of N-methylpiperidine with \textit{n}-hexene, as investigated by \citet{Liu2017} and \citet{Iron2019}. In the full catalytic cycle, the amine-coordinated $\eta^2$-azametallacyclic complex serves as the active species, which undergoes ligand exchange with $n$-hexene to form a weakly bound $\pi$-complex. The subsequent insertion of the C=C bond into the Sc–C bond proceeds via a four-center transition state, yielding a ring-expanded azametallacyclic intermediate. This transformation constitutes the key C–C bond-forming step preceding C–H activation and regeneration of the catalyst. This reaction provides a stringent benchmark because it features a single, well-defined four-center transition state on a shallow potential surface.

Application of the native and low-level FSM workflows with UMA-M model to this system led to different outcomes. The native FSM TS guess failed to converge to a saddle point within 250 optimization cycles. In contrast, the low-level-refined guess converged to the correct transition state after only 22 optimization cycles. The initial $\omega$B97X-V/def2-TZVP frequency calculation based on the native guess revealed five imaginary modes, with the largest at $308\text{i cm}^{-1}$, approximately 18~\kcalmol\ above the reference TS, and a root-mean-square deviation (RMSD) of 0.75 ~\AA. In contrast, the low-level-refined guess displayed a largest imaginary mode of ($243\text{i cm}^{-1}$), consistent with the reference value ($228\text{i cm}^{-1}$), and was approximately 3~\kcalmol\ higher in energy and RMSD of 0.63 ~\AA relative to the reference structure.
Both MACE-OMol25 workflows successfully located the reference transition state, though with different computational costs. The native MACE-OMol25 guess exhibited a largest imaginary mode of $282\text{i cm}^{-1}$, an error of 19~\kcalmol, and RMSD of 0.76~\AA, converging to the reference TS after 201 optimization steps. The corresponding low-level-refined guess improved these metrics ($231\text{i cm}^{-1}$, 3~\kcalmol, and 0.66 ~\AA) and converged to the correct transition state in 20 optimization steps. 

As expected, the native guesses from the FSM initially overestimated the barrier due to the truncated path optimization. The UMA-M model predicted a barrier height of 10.9~\kcalmol\ and a reaction energy of $-$14.9~\kcalmol\ when evaluated on the low-level-refined TS geometry, overestimating the barrier by by 5.4~\kcalmol\ relative to the DLPNO-CCSD(T)/CBS reference (5.5 and $-$16.9~\kcalmol) while reproducing the reaction energy within 2.0~\kcalmol. The MACE-OMol25 model showed larger barrier errors (13.9~\kcalmol, an overestimation of 8.4~\kcalmol) but predicted the reaction energy within 0.4~\kcalmol\ of the DLPNO-CCSD(T)/CBS value ($-$17.3 vs $-$16.9~\kcalmol). Both models overestimate the barrier relative to $\omega$B97X-V/def2-TZVP as well, indicating systematic errors in the MLIP energetics near the saddle point region. Importantly, despite these energetic discrepancies, the low-level-refined geometries were of sufficient quality for the high-level P-RFO to converge to the correct transition state in 20–22 optimization steps.

\subsubsection{Iron-Catalyzed Transfer Hydrogenation} 
The iron-catalyzed transfer hydrogenation of acetophenone with isopropanol, mediated by Fe(II) PNNP eneamido complexes,\cite{Prokopchuk2012} is a prototypical example of base-metal catalysis. The key elementary step involves inner-sphere proton-transfer, which has been extensively characterized theoretically and included in the MOBH35 dataset, providing high accuracy DLPNO-CCSD(T)/CBS calculations of the activation and reaction energy. \cite{Iron2019} 

Application of the FSM workflows with the UMA-M and MACE-OMol25 successfully located the reference TS in all four cases. The native guesses converged in 30 and 28 gradient evaluations, respectively, while the low-level refinement reduced those to 11 and 14 gradient evaluations. 

As expected, the native FSM guesses substantially overestimate the activation barrier due to the truncated path optimization, predicting barriers of 53.0 and 52.2~\kcalmol\ for UMA-M and MACE-OMol25 respectively, while the reference DLPNO-CCSD(T)/CBS value is 16.4~\kcalmol. Low-level refinement dramatically improves these estimates: the UMA-M refined barrier of 15.3~\kcalmol\ is within 1.1~\kcalmol\ of the DLPNO-CCSD(T)/CBS value and 0.3~\kcalmol\ of the $\omega$B97X-V/def2-TZVP result (15.0~\kcalmol), while the MACE-OMol25 refined barrier of 13.1~\kcalmol\ underestimates the reference by 3.3~\kcalmol. For the reaction energy, UMA-M predicts 1.9~\kcalmol, in exact agreement with the DLPNO-CCSD(T)/CBS value and within 0.5~\kcalmol\ of the $\omega$B97X-V/def2-TZVP (1.4~\kcalmol). MACE-OMol25 predicts $-$0.4~\kcalmol, incorrectly reversing the sign of the reaction energy. The UMA-M and MACE-OMol25 refined structures exhibit $\omega$B97X-V/def2-TZVP imaginary frequencies of $1352\text{i}$/$64\text{i cm}^{-1}$ (RMSD = 0.395~\AA) and $1382\text{i}$/$64\text{i cm}^{-1}$ (RMSD of 0.388~\AA), respectively, compared to a single mode at $1387\text{i cm}^{-1}$ for the reference transition state, confirming that both models closely reproduce the saddle-point topology of the DFT surface. The error of the UMA-M model on this system is consistent with the error that might be observed from changes in level of theory between $\omega$B97X-V/def2-TZVP and $\omega$B97M-V/def2-TZVPD (OMol25). The accuracy of the UMA-M refined barrier on the MLIP-surface alone is notable and suggests that for this reaction, reliable barrier estimates can be obtained directly from the MLIP without recourse to DFT single point correction. 

\subsubsection{Platinum-Mediated Metallabenzene Re-arrangement}
The platinum-mediated metallabenzene rearrangement is an intramolecular C–C coupling step that converts a metallabenzene complex to its cyclopentadienyl (Cp) analogue, as investigated by \citet{Iron2003} and included in the MOBH35 benchmark set. \cite{Iron2019} In this transformation, the aromatic metallabenzene precursor undergoes direct C–C bond formation through a three-centered transition state characterized by an asymmetric M{=}C{-}C{=}C{-}C{=}C bonding motif. The reaction proceeds via carbene migratory-insertion-type mechanism, yielding an $\eta^5$–Cp product. This system represents a demanding benchmark as it involves contraction of an aromatic six-membered metallacyclic ring to a five-membered Cp ligand through direct C–C coupling. 

Application of the FSM with UMA-M produced guess structures characterized by largest initial imaginary modes of $706\text{i}$ and $405\text{i cm}^{-1}$ for the native and low-level-refined workflows, respectively, compared to the reference transition-state frequency of $379\text{i cm}^{-1}$. The native guess structure was approximately 30~\kcalmol~ higher in energy than the reference TS and had an RMSD of 0.65 ~\AA, while the low-level-refined guess was only 0.2~\kcalmol\ higher in energy and had an RMSD of 0.39 ~\AA. Both guesses converged to the reference saddle point, requiring 91 and 22 optimization steps, respectively.

The MACE-OMol25 model showed larger initial deviations. The native guess exhibited an imaginary frequency of $1305\text{i cm}^{-1}$, lay 37~\kcalmol\ above the reference, and an RMSD of 0.66 ~\AA. This converged to the reference saddle point in 97 optimization steps, though the converged structure retained a second spurious imaginary frequency of $24\text{i cm}^{-1}$. The low-level-refined guess, when evaluated with $\omega$B97X-V/def2-TZVP, exhibited a largest imaginary frequency of $429\text{i cm}^{-1}$, was approximately 1~\kcalmol\ higher in energy than the reference transition state, and had an RMSD of 0.39 ~\AA. Despite these initial metrics, the subsequent high-level P-RFO optimization failed to converge within 250 steps, with the final frequency analysis revealing no imaginary modes indicating that the optimization converged to a local minimum. 

Single-point $\omega$B97X-V/def2-TZVP energy calculations on the refined UMA-M and MACE-OMol25 geometries predicted barrier heights of 21.9 and 23.2~\kcalmol, respectively, in close agreement with the DLPNO-CCSD(T)/CBS reference (19.0~\kcalmol) and $\omega$B97X-V/def2-TZVP optimized value (21.2~\kcalmol). The predicted reaction energies ($-$40.5 and $-$43.2~\kcalmol) are similarly consistent with the DLPNO-CCSD(T)/CBS ($-$45.9~\kcalmol) and $\omega$B97X-V/def2-TZVP ($-$42.4~\kcalmol) references. Similar to prior sections, the results on this test case show that MLIP-refined geometries are of high accuracy.

\subsubsection{Rhodium-Catalyzed Oxidative Carbonylation of Toluene}

\begin{figure*}
\begin{center}
\includegraphics[width=0.9\textwidth,trim=12cm 13cm 17cm 10cm, clip]{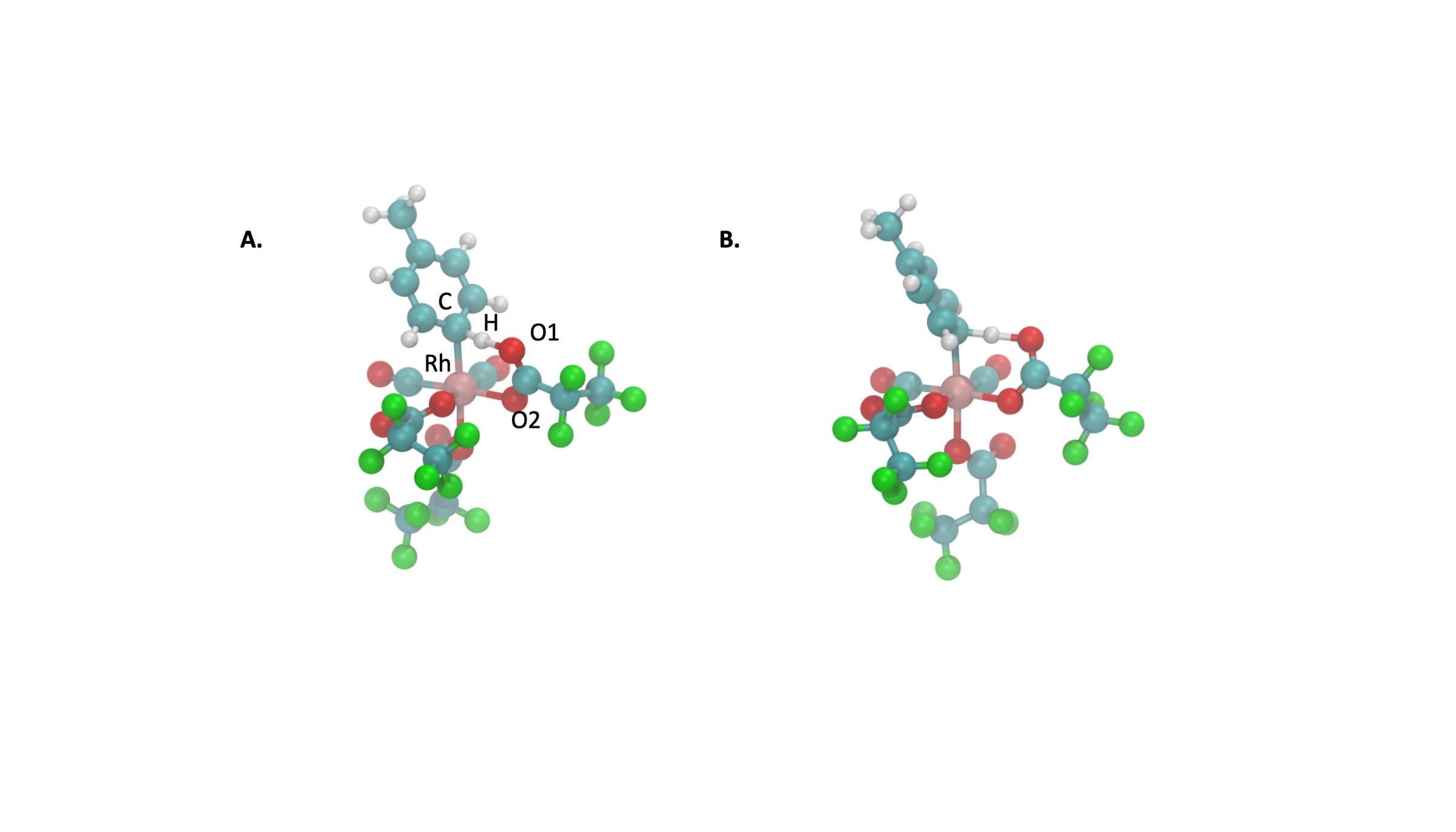}
\caption{Transition-state structures for the Rh(III)-catalyzed C–H activation step in oxidative carbonylation of toluene. (A) Reference transition state optimized at the $\omega$B97X-V/def2-TZVP level of theory. (B) Transition state obtained from the UMA-M low-level-refined workflow after DFT refinement. Key atoms defining the reactive core are labeled in panel (A).}
\label{fig:Rh_TS}
\end{center}
\end{figure*}

The Rh(III)-catalyzed oxidative carbonylation of toluene to \textit{p}-tolouic acid in trifluoroacetic acid represents a well-characterized example of transition metal-mediated aromatic C–H activation. Combined experimental and theoretical investigations by \citet{Zakzeski2009} established that the catalytically active species is $\text{Rh(CO)}_2\text{(TFA)}_3$, where $\text{TFA}^-$ denotes triflouroacetate anions. The proposed mechanism proceeds through reversible coordination of toluene via the para position, followed by rate limiting electrophilic C–H bond activation at the para carbon. Theoretical analysis at B3LYP/6-311G**//B3LYP/6-31G* level of theory confirmed that C–H activation proceeds through an electrophilic substitution mechanism rather than via agostic C–H interactions, as evidenced by Rh–H distances exceeding 2.4~\AA\ in both the coordinated complex and transition state.\cite{Zakzeski2009} We select the rate-limiting C–H activation elementary step as our test case, using the $\text{Rh(CO)}_2\text{(TFA)}_3\text{(C}_6\text{H}_5\text{CH}_3\text{)}$ complex as the reactant and the corresponding $\text{Rh(CO)}_2\text{(TFA)}_2\text{(TFAH)}\text{(C}_6\text{H}_4\text{CH}_3\text{)}$ species as the product. This system provides a demanding benchmark because it requires accurate treatment of an octahedral d$^6$ Rh(III) center with multiple carboxylate ligands. 

Application of the FSM with UMA-M and MACE-OMol25 potentials successfully located the C–H activation transition state in all four workflow combinations, though the converged structures corresponded to two distinct conformers distinguished by the orientation of the trifluoroacetate ligands shown in Figure \ref{fig:Rh_TS}. The native work flows for both models converged in 82 optimization steps each, yielding transition-state structures with largest imaginary frequencies of $885\text{i}$ (Figure \ref{fig:Rh_TS}B) and $890\text{i cm}^{-1}$ for UMA-M and MACE-OMol25 respectively, compared to the reference value of $831\text{i cm}^{-1}$ (Figure \ref{fig:Rh_TS}A. The low-level-refined workflows converged to a second conformer characterized by larger imaginary frequencies of $937\text{i}$ and $938\text{i cm}^{-1}$ for UMA-M and MACE-OMol25, respectively, requiring 55 and 76 optimization steps. Both conformers correspond to the same electrophilic C–H activation mechanism and differ only in the rotational orientation of the side groups. This distinction does not affect the mechanistic interpretation, as the reaction does not involve stereo selectivity at this elementary step. 

The energetic analysis reveals modest deviations between the converged conformers. Single-point energies of TS conformers resulting from the native-workflow lie approximately 1.6~\kcalmol\ above the reference saddle point at the $\omega$B97X-V/def2-TZVP level of theory, while low-level-refined structures are approximately 1.4~\kcalmol\ higher than the reference saddle point. These small energy differences fall within the range expected for ligand rotational isomers on a relatively flat PES. 

To assess the structural quality of the converged transition states, we compare key geometric parameters of the reactive core defined by five atoms directly involved in the bond breaking and formation (H, C, O1, O2, and Rh; see Figure \ref{fig:Rh_TS}A across all workflows and against both the $\omega$B97X-V/def2-TZVP reference and B3LYP/6-311G**//B3LYP/6-31G* values reported by \citet{Zakzeski2009} (Table~\ref{tab:rh_geometries}). The native FSM guesses from both models show substantially elongated C–H bonds (1.55-1.56~\AA\ vs 1.28~\AA\ reference) and Rh–H distances (2.71–2.72\AA\ vs. 2.31~\AA), reflecting the incomplete optimization of the breaking C–H bond at the FSM guess stage. The H–O1 distance is similarly overestimated at 1.47~\AA\ compared to 1.37~\AA. In contrast, the metal–carbon (Rh–C) and metal–oxygen (Rh–O2) distances are well reproduced even in the native guesses (within 0.01-0.02~\AA\ of reference), indicating that the FSM and UMA-M/MACEOMol potentials capture the metal coordination geometry of the transition state prior to any refinement. 

Following low-level refinement, both UMA-M and MACE-OMol25 yield reactive core geometries in excellent agreement with $\omega$B97X-V/def2-TZVP reference. The C–H distances (1.30 and 1.31~\AA) and Rh–H distances (both 2.33~\AA) are within 0.03~\AA\ of the reference values, while the H–O1 bond lengths (both 1.36~\AA) are within 0.01~\AA. The Rh–C and Rh–O2 distances are similarly well reproduced, deviating by at most 0.04~\AA\ across both models. This level of agreement demonstrates that the MLIP PESs themselves accurately minimize transition-state geometries without recourse to DFT. Notably, the Rh–H distances remain above 2.3~\AA\ in all refined structures, consistent with the electrophilic substitution mechanism characterized by \citet{Zakzeski2009} and confirming the MLIPs correctly reproduce the non-agostic nature of this transition state. Collectively, these geometric comparisons confirm that both MLIPs faithfully reproduce geometries of the reactive core consistent with DFT PESs, and that the remaining energetic discrepancies originate from the conformational degrees of freedom peripheral to the reaction center. 

\begin{table}[ht]
    \centering
    \caption{Geometric parameters of the reactive core at the transition state as calculated by various potentials.}
    \label{tab:rh_geometries}
    \begin{tabular}{l ccccc}
        \toprule
         Method & H–O$_1$ & C–H & Rh–H & Rh–C & Rh–O$_2$\\
        \midrule
        Literature\cite{Zakzeski2009} & 1.38 & 1.28 & 2.40 & 2.17 & 2.11 \\
        $\omega$B97X-V/def2-TZVP & 1.37 & 1.28 & 2.31 & 2.18 & 2.05 \\
        UMA-M Native & 1.47 & 1.55 & 2.71 & 2.17 & 2.06 \\
        UMA-M Refined & 1.36 & 1.30 & 2.33 & 2.14 & 2.06 \\
        MACE-OMol25 Native & 1.47 & 1.56 & 2.72 & 2.19 & 2.06 \\
        MACE-OMol25 Refined & 1.36 & 1.31 & 2.33 & 2.16 & 2.07 \\
    
        \bottomrule
    \end{tabular}
\end{table}

The near-identical reactive-core geometries across both conformers and both models further clarify the origin of modest energetic deviations ($\sim$1.4–1.6~\kcalmol) observed between the converged transition states and the reference saddle point. Because the bond-breaking and bond-forming geometric parameters are closely reproduced, the energetic differences can be attributed to differing orientations of the peripheral TFA ligands rather than errors in description of the reaction mechanism. These conformational differences arise from a combination of the initial guess geometry provided to the P-RFO optimizer and the optimizers trajectory across the saddle point region; on a relatively flat surface with respect to ligand rotation, small differences in the starting geometry can direct convergence toward distinct but mechanistically equivalent conformers. 
This underscores an important practical consideration for transition-state searches on large organometallic systems: peripheral-ligand flexibility can produce multiple valid saddle points with nearly degenerate energies, and the lowest-energy conformer is not guaranteed to be found without systematic conformational sampling.

\section{Discussion}\label{sec:discussion}
The systematic benchmarking of algorithm–MLIP combinations across organic and organometallic reactions reveals several key insights into the design of efficient hybrid workflows for automated TS searches. Most fundamentally, the choice of reaction-path algorithm has a more profound impact on search reliability than the choice of MLIP. The FSM achieved success rates of 88.9\% and 90.3\% on the Baker and Sharada sets, respectively, compared to 70.8\% and 62.0\% for the CI-NEB method on the Baker and Sharada sets, respectively (Appendix~A), when averaged over all six potentials. This performance gap persisted even after low-level refinement, indicating that the algorithmic differences in how initial guesses are generated, rather than their subsequent local optimization, largely determines whether the downstream DFT P-RFO converges to the correct saddle point. In cases where the CI-NEB method successfully located a viable path, the resulting TS search converged quickly and yielded TS geometries and barrier heights comparable in accuracy to those obtained via FSM. The relative performance of the CI-NEB method may be strongly implementation dependent, and improved interpolation or optimization heuristics could reduce the observed gap in performance.

Among the MLIPs evaluated, the systematic advantage of OMol25-trained models (MACE-OMol25, UMA-M, UMA-S, eSEN-S) over alternative potentials (GFN2-xTB, AIMNet2) is striking. Across 29 unique reactions from the Baker and Sharada sets, the OMol25 models were successful in 91.8\% of searches with an average cost of 11.7 DFT gradient evaluations compared to an 84.5\% success rate and 15.5 gradient evaluations for non-OMol25 potentials. Aside from neural-network architectural differences, an advantage arises from the quantity and composition of the training data, the OMol25 dataset contains approximately 100M structures, including reactive snapshots, transition-state region configurations, and systematically diverse conformers calculated at the $\omega\text{B97M-V/def2-TZVPD}$ level of theory. Models such as UMA-M and UMA-S further benefit from inclusion of additional data sources \cite{Chanussot2021,Sriram2022,Tran2023,Barroso-Luque2024,Levine2025} spanning broader chemical domains, which appeared to enhance their transferability to transition metal complexes and supported catalytic systems. 

The value of low-level refinement on the MLIP surface is evident across all chemical regimes studied. For the organic benchmarks, low-level refinement reduced the average number of DFT gradient evaluations from 14.7 to 5.0 on the Baker set and from 36.4 to 8.0 on the Sharada set, with reductions of roughly 50\% in RMSD between native and refined guess structures. On the Poly25 closed-shell-polymerization benchmark, refinement reduced UMA-M and MACE-OMol25 costs from 31.5 and 22.0 to 5.0 and 5.2 gradient evaluations, respectively. For transition-metal systems, similar cost reductions were observed alongside substantial improvements in guess quality. This demonstrates that inexpensive local optimization on the MLIP surface provides geometries that are significantly closer to the target saddle point and should be a routine component of hybrid transition-state-search workflows. 

Comparing model performance across chemical regimes, MACE-OMol25 performed best for small organic systems, requiring an average of only 3.8 gradient evaluations across the Baker and Sharada sets with only a single failure, achieving a comparable cost of 5.2 gradient evaluations on the Poly25 set where the sole failure was attributable to the FSM interpolation rather than the potential. In contrast, UMA-M showed superior performance for transition-metal reactions, including both the near-in-distribution MOBH35 systems and the out-of-distribution rhodium-catalyzed C–H activation case. This improvement reflects UMA-M's broader training data, which includes metal-surface and metal-adsorbate configurations from the OC20\cite{Chanussot2021} and OC22\cite{Tran2023} in addition to the OMol25 molecular dataset, exposing the model to a wider range of metal-ligand coordination environments and oxidation states.  

Taken together, these results indicate that MLIP-accelerated transition-state searches based on algorithms such as the FSM with low-level refinement workflow are already suitable for high throughput studies. For small organic reactions, these models achieve near-DFT reliability at a fraction of the cost. For more complex transition-metal systems, UMA-M demonstrated promising transferability, but it should still be applied with caution, as capturing the complexities of spin or charge re-organization along reaction coordinates remains a difficult task for current algorithms and models. For well-covered areas of chemical space such as small organic and closed-shell polymerization reactions, the demonstrated reliability and low cost of these workflows made them suitable for high-throughput screening, where approximate barrier heights can be obtained rapidly from MLIP-refined geometries with DFT single-point corrections applied as needed. For transition metal systems, MLIP-optimization substantially reduces the initial cost of mechanistic exploration, but DFT validation of the final saddle point remains essential given the higher error observed in select cases. 

\subsection{Current limitations/Failure Mode analysis}

\subsubsection{Failure Modes Introduced by Low-Level Refinement}
\label{sec:sella_failures}

\begin{figure*}
\includegraphics[width=1.0\textwidth]{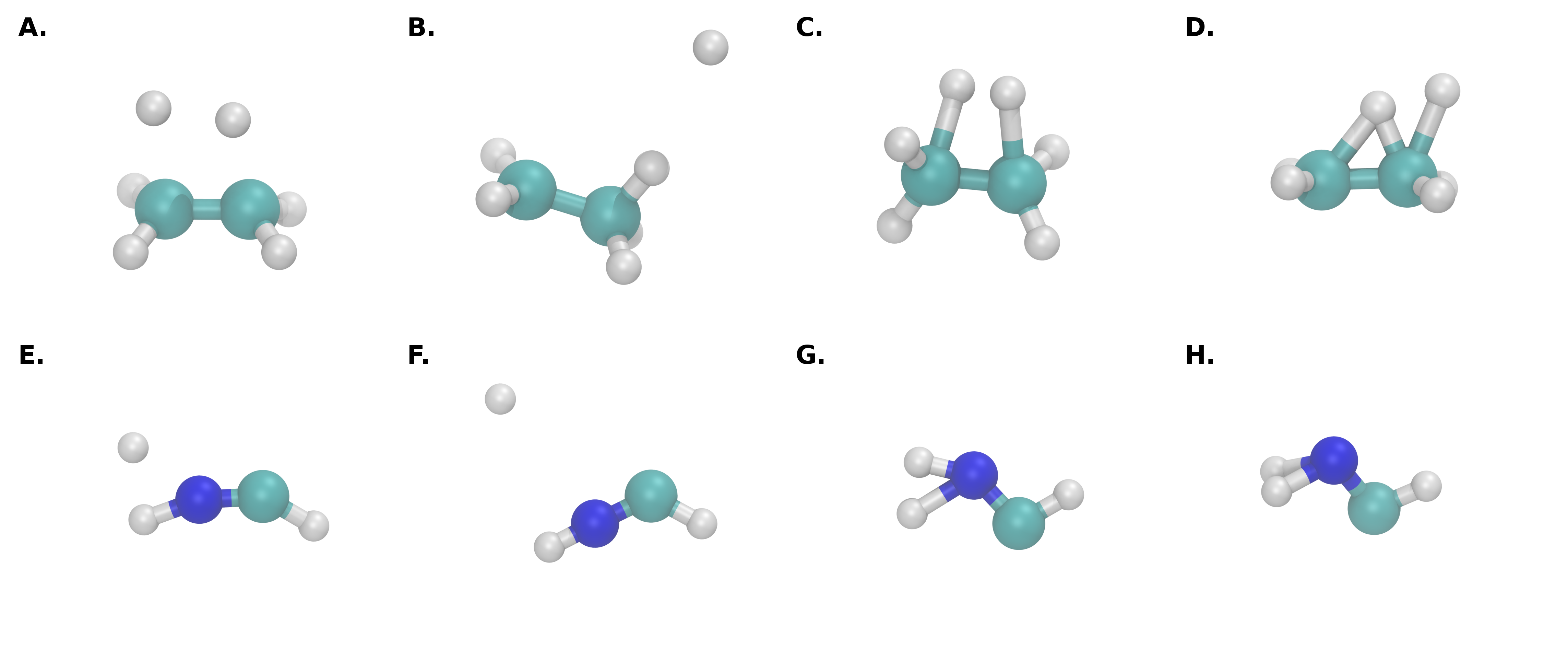}
\caption{Native guesses, low-level refined structures, and corresponding converged $\omega$B97X-V/def2-TZVP transition states for reactions 11 (CH$_3$CH$_3$ $\rightarrow$ CH$_2$CH$_2$ + H$_2$) and 24 (HCNH$_2$ $\rightarrow$ HCN + H$_2$) of the Baker set using FSM with UMA-M. (A) Native MLIP guess for reaction 11. (B) Low-level-refined MLIP structure for reaction 11. (C) Higher-energy saddle point obtained from (B) after DFT refinement. (D) Reference transition state for reaction 11 at the $\omega$B97X-V/def2-TZVP level. (E) Native MLIP guess for reaction 24. (F) Low-level-refined MLIP structure for reaction 24. (G) Higher-energy second-order saddle point obtained from (F) after DFT refinement. (H) Reference transition state for reaction 24 at the $\omega$B97X-V/def2-TZVP level.}
\label{fig:HCNH2_ETHDEHYD}
\end{figure*}

Several of the observed failures represent near-miss cases, such as the low-level-refined UMA-S and UMA-M searches on reaction 11, (ethane dehydrogenation). The native guesses from UMA-S and UMA-M (Shown in Figure~\ref{fig:HCNH2_ETHDEHYD}A) converge to the reference saddle point (Figure~\ref{fig:HCNH2_ETHDEHYD}D) which has an imaginary frequency of $2267\text{i cm}^{-1}$. Low-level refinement on the UMA-S and UMA-M surfaces results in the structure shown in Figure~\ref{fig:HCNH2_ETHDEHYD}B, which when input to the high-level DFT refinement, results in convergence to a saddle point that lies approximately 8~\kcalmol\ higher in energy, and has an imaginary frequency of $2130\text{i cm}^{-1}$ shown in \ref{fig:HCNH2_ETHDEHYD}C. IRC calculations at the $\omega$B97X-V/def2-TZVP level of theory validated that this higher energy saddle point corresponds to correct transformation via an alternate mechanism; symmetric dissociation of the Hydrogen atoms. In contrast, the lower energy, reference saddle point undergoes an asymmetric dissociation of the Hydrogen atoms. 

A similar near-miss occurred for the UMA-M refined search on reaction 16 (Claisen rearrangement). The located saddle-point had an imaginary frequency of $532\text{i cm}^{-1}$ and was 43~\kcalmol\ higher in energy than the reference saddle point, which had an imaginary frequency of $633\text{i cm}^{-1}$. IRC calculations reveal the higher energy saddle point corresponds to a similar transformation wherein both reactant and product states are in a higher-energy cis configuration. In many other cases the off-target saddle points found correspond to chemically-distinct transformations.

For a native search to succeed, the FSM must generate a geometry whose largest imaginary mode corresponds to the transformation of interest. Low-level refinement on the MLIP or SE-DFT potential introduces additional error to this process in two ways. First, Sella constructs an approximate Hessian that deviates from the exact one, even when derived from DFT gradient evaluations. Second, when that approximate Hessian is computed from an MLIP, model specific curvature errors can distort the relative ordering of the imaginary modes. In test cases such as reaction 11 of the Baker set (ethane dehydrogenation), where several first-order saddle points are geometrically similar, this misranking redirects the optimizer toward an incorrect transition state. To isolate the source of error, refinement using the Sella Python package was repeated using $\omega$B97X-V/def2-TZVP gradients on the same FSM UMA-M native guess. The subsequent high-level refinement in Q-Chem successfully located the reference saddle point in two optimization cycles, confirming that the failure arises from Hessian inaccuracy in the UMA-M model rather than from the refinement algorithm itself. Diagonalization of the UMA-M finite-difference Hessian for the native guess yielded imaginary frequencies of $3969\text{i}$, $949\text{i}$, and $762\text{i cm}^{-1}$, compared to $2343\text{i}$, $843\text{i}$, and $598\text{i cm}^{-1}$ at $\omega$B97X-V/def2-TZVP level of theory demonstrating substantial curvature distortion at the guess geometry that drives convergence toward an alternate saddle point. 

A second systematic failure was observed for reaction 24 of the Baker set, (HCNH$_2$ $\rightarrow$ HCN + H$_2$). All native searches with OMol25 trained models located the reference saddle point (Figure \ref{fig:HCNH2_ETHDEHYD}H, whereas low-level-refined guesses (\ref{fig:HCNH2_ETHDEHYD}F) converged to the same higher-energy second-order saddle point (Figure \ref{fig:HCNH2_ETHDEHYD}A). Because eSEN-S, UMA-M, UMA-S, MACE-OMol25 all produced identical outcomes, the MACE-OMol25 run serves as a representative search. At the $\omega$B97X-V/def2-TZVP level of theory, the native guess (\ref{fig:HCNH2_ETHDEHYD}E) exhibited a single imaginary mode of $2075\text{i cm}^{-1}$, while the MACE-OMol25 model predicted two modes of $1995\text{i}$ and $500\text{i cm}^{-1}$. The presence of this secondary imaginary mode indicates significant Hessian error: the RS-P-RFO procedure attempts to maximize the largest mode while minimizing all others, and the secondary mode skews the step direction, leading to convergence on a higher-energy second-order saddle point (Figure \ref{fig:HCNH2_ETHDEHYD}G). This failure also arises due to the use of approximate Hessian update heuristics. The P-RFO optimizer as implemented in Q-Chem employs the Powell–Murtagh–Sargent update scheme to approximate the Cartesian Hessian at new geometries as optimization proceeds. \cite{murtagh1970computational,powell1971recent} In this approximate Hessian, the secondary mode is eliminated during optimization and once the optimization convergence criteria are satisfied, the optimization terminates. A final full Hessian calculation at the optimized geometry restores the missing mode and reveals the error in the approximate Hessian. This behavior underscores the sensitivity of the P-RFO algorithm to inaccuracies in the Hessian due to either approximation or underlying model error. 

\subsubsection{General failure modes}
\label{sec:general_failure}
Beyond the refinement-specific failures analyzed above, several additional failure mechanisms were observed that are not attributable to the low-level refinement step. The most common involved convergence to a local minimum rather than the intended first-order saddle point, typically occurring when the initial guess lay outside the basin of attraction of the target transition state. A related behavior was convergence to alternate first-order saddle points corresponding to different chemical transformations, as observed in reaction 5 (cyclopropyl ring-opening) and reaction 6 (bicyclo[1.1.0]butane rearrangement) of the Baker set. A smaller subset of failures originated from numerical issues during Q-Chem P-RFO refinement itself. These included self-consistent field (SCF) convergence failures, internal-coordinate breakdowns when the delocalized coordinate system became ill-defined (as observed for reaction 22 of the Baker set with CI-NEB and MACE-OMol25), and fatal optimization errors in the P-RFO step, typically failure to determine appropriate step size. 

\subsection{Literature comparison}

To contextualize the efficiency gains afforded by the MLIP-accelerated workflow, we compared our results directly to previous benchmarks of the FSM with redundant-internal-coordinate interpolation that used DFT for both the reaction-path finding and the subsequent TS refinement.\cite{Marks2024incorporation} This comparison is particularly informative because both studies employ identical reactant and product geometries for the Baker and Sharada benchmark sets and use the same $\omega$B97X-V/def2-TZVP level of theory for P-RFO TS refinement. In the all-DFT FSM workflow, transition-state searches on the Baker set required an average of 73 total gradient evaluations (59 for the FSM and 14 for the P-RFO) with 100\% success rate, while the Sharada set required 90 gradient evaluations (60 for the FSM and 30 for the P-RFO), also with a 100\% success rate. The key distinction in the present work is that the MLIP bears the entire computational burden of the reaction path finding step, with DFT only employed only for the final P-RFO refinement. 

Incorporation of low-level refinement on the MLIP surface dramatically reduced the overall DFT cost of the search. MACE-OMol25-refined searches achieved a 95.8\% success rate and required only 2.9 gradient evaluations on Baker, a 96\% reduction relative to the full DFT search and a 79\% reduction in required DFT gradient calls during high-level TS refinement alone. Similar improvements were observed on the Sharada set; MACE-OMol25 maintains a 100\% success rate while averaging 5.7 gradient evaluations, a 94\% and 81\% reduction relative to the full DFT searches and DFT high-level TS refinement step respectively. Notably, the DFT costs of native workflows are comparable to the P-RFO refinement step alone in the all-DFT searches, demonstrating that replacing DFT FSM calculations with MLIP evaluations eliminates the majority of the computational expense while incurring only a minor loss in reliability.

Recent work by \citet{Hait2025} proposed an alternative MLIP-based approach using geodesic path construction on the eSEN-S-sm-cons MLIP surface.\cite{Hait2025} Their method globally optimizes densely discretized paths (typically $\sim$40 nodes), requiring hundreds of MLIP gradient evaluations before final P-RFO refinement in Q-Chem. Applied to the Sharada benchmark, this approach successfully generated viable transition-state guesses for all nine reactions, all of which converged to reference transition states in $\sim$30\% fewer P-RFO iterations than those from DFT-FSM calculations run natively in Q-Chem. In comparison, we find that the MACE-OMol25 model outperforms UMA and eSEN-S models on small organic systems, achieving a 100\% success rate on the Sharada set when performing local refinement with the MACE-OMol25 surface, at an average cost of 5.7 gradient evaluations, compared to the average of 16.7 for the geodesic method. The geodesic framework would likely benefit from integration with MACE-OMol25, as the model's improved accuracy across organic systems could further reduce computational cost. 

A practical consideration concerns the assumption that MLIP gradients are computationally negligible. While reasonable for small systems and moderate-scale studies, this assumption weakens for larger systems or high-throughput campaigns. The more computationally intensive geodesic optimization appears to provide improved reliability for challenging cases, where truncated path optimization of the FSM approaches may lead to failed TS searches. Both frameworks clearly demonstrate that well-trained MLIPs can produce transition-state guesses of sufficient accuracy to reduce DFT cost by one to two orders of magnitude relative to conventional \textit{ab initio} workflows.

\subsection{Future Directions for Improvement}
Several of the observed error modes originated from errors in the local curvature information. Improved Hessian information offers a direct path forward. The computational efficiency of MLIPs makes it practical to evaluate full Hessians at each optimization step for methods such as RS-P-RFO and P-RFO, which has been shown to improve convergence reliability and efficiency.\cite{yuan2024analytical} Emerging techniques for rapid Hessian evaluation through automatic differentiation promise to make these approaches model-agnostic and inexpensive.\cite{burger2025shoot} Their success, however, depends on the underlying MLIP producing physically meaningful second derivatives.

A complementary direction is the development of reaction path-finding algorithms and MLIPs capable of handling variable spin and charge states along the reaction coordinate. Many catalytic and redox processes such as the catalyzed oxidation of methanol on titania-supported oxide catalysts\cite{Bronkema2007,Goodrow2008} involve transitions between electronic states that require multi-reference treatment and cannot be described on a single PES. Extending MLIP-based workflows to such systems will require both models accurate across electronic states and reaction-path methods capable of switching between electronic states. 

\section{Conclusions}
This work provides a systematic benchmarking of modern MLIPs and demonstrates that, when coupled with appropriate reaction-path and refinement algorithms, these models enable reliable and efficient automated TS searches across broad areas of chemical space at a fraction of the computational cost of DFT. Through systematic benchmarking of 24 algorithm–potential combinations across 58 diverse reactions, we demonstrate that the choice of reaction-path algorithm fundamentally determines search reliability: the FSM achieved 88–90\% success rates compared to 63–71\% for the CI-NEB implementation tested here. Among MLIPs, models trained on the OMol25 dataset (MACE-OMol25, UMA-M, UMA-S, eSEN-S) systematically outperformed alternatives, achieving 91.8\% success compared to 84.5\% for non-OMol25 potentials.

For small organic systems, MACE-OMol25 combined with low-level refinement provides optimal performance, locating reference transition states in 96.6\% of attempts while requiring an average of only 3.8 DFT gradient evaluations, a 94–96\% reduction compared to conventional all-DFT workflows. For transition metal containing systems, UMA-M demonstrated superior transferability, successfully handling both near-distribution test cases drawn from the MOBH35 benchmarks and the out-of-distribution rhodium-catalyzed C–H activation system.

Low-level refinement on the MLIP surface before high-level DFT optimization emerged as a critical workflow component, reducing average DFT costs by approximately three fold, while maintaining success rates. For organic systems, the modest reliability decrease observed with low-level refinement arose from MLIP curvature errors that can redirect optimization toward alternate saddle points, particularly when multiple geometrically similar saddle points exist. For transition-metal systems, the primary challenges were convergence during high-level refinement and sensitivity to the quality of the initial-guess geometry, reflecting the greater complexity of these systems. 

These results demonstrate that MLIP-accelerated transition-state searches have matured to the point of practical deployment for high-throughput mechanistic studies. The workflows achieve near-DFT reliability for organic and closed-shell polymerization reactions while enabling orders-of-magnitude speedups. For transition-metal systems, current models show promise but require continued development. Given the demonstrated reliability and negligible computational cost of MLIP-based reaction-path finding relative to DFT, we recommend that MLIP pre-optimization with models such as MACE-OMol25 and UMA-M be adopted as the default first stage in transition-state-search workflows, replacing semi-empirical or low-level DFT methods for this purpose. The demonstrated efficiency and scalability of these hybrid workflows opens new possibilities for computational exploration of complex reaction networks in catalysis, materials synthesis, and synthetic chemistry. 

\section*{Author contributions}
\textbf{Jonah Marks}: Conceptualization, methodology, software, investigation, formal analysis, data curation, writing -- original draft, visualization, editing. 
\textbf{Jonathon Vandezande}: Methodology, resources (Poly25 dataset) writing -- review \& editing.
\textbf{Joseph Gomes}: Conceptualization, methodology, supervision, funding acquisition, resources, writing -- review \& editing. 

\section*{Conflicts of interest}
J.E.V. is an employee of the Rowan Scientific Corporation, which is commercializing the ML-FSM code used in this paper. All other authors declare no competing interests.

\section*{Data availability}
The data supporting this study, including input geometries and scripts used to perform the transition state benchmarking calculations, will be made publicly available in a GitHub repository upon publication of this work and archived with a DOI.

\section*{Acknowledgements}
J.G acknowledges start up funding from The University of Iowa. This research was supported in part through computational resources provided by The University of Iowa.



\balance

\renewcommand\refname{References}

\bibliography{rsc} 

\providecommand*{\mcitethebibliography}{\thebibliography}
\csname @ifundefined\endcsname{endmcitethebibliography}
{\let\endmcitethebibliography\endthebibliography}{}
\begin{mcitethebibliography}{64}
\providecommand*{\natexlab}[1]{#1}
\providecommand*{\mciteSetBstSublistMode}[1]{}
\providecommand*{\mciteSetBstMaxWidthForm}[2]{}
\providecommand*{\mciteBstWouldAddEndPuncttrue}
  {\def\EndOfBibitem{\unskip.}}
\providecommand*{\mciteBstWouldAddEndPunctfalse}
  {\let\EndOfBibitem\relax}
\providecommand*{\mciteSetBstMidEndSepPunct}[3]{}
\providecommand*{\mciteSetBstSublistLabelBeginEnd}[3]{}
\providecommand*{\EndOfBibitem}{}
\mciteSetBstSublistMode{f}
\mciteSetBstMaxWidthForm{subitem}
{(\emph{\alph{mcitesubitemcount}})}
\mciteSetBstSublistLabelBeginEnd{\mcitemaxwidthsubitemform\space}
{\relax}{\relax}

\bibitem[Grambow \emph{et~al.}(2020)Grambow, Pattanaik, and Green]{grambow2020deep}
C.~A. Grambow, L.~Pattanaik and W.~H. Green, \emph{J. Phys. Chem. Lett.}, 2020, \textbf{11}, 2992--2997\relax
\mciteBstWouldAddEndPuncttrue
\mciteSetBstMidEndSepPunct{\mcitedefaultmidpunct}
{\mcitedefaultendpunct}{\mcitedefaultseppunct}\relax
\EndOfBibitem
\bibitem[Makos \emph{et~al.}(2021)Makos, Verma, and Larson]{Makos2021}
M.~Z. Makos, N.~Verma and E.~C. Larson, \emph{J. Chem. Phys.}, 2021,  024116\relax
\mciteBstWouldAddEndPuncttrue
\mciteSetBstMidEndSepPunct{\mcitedefaultmidpunct}
{\mcitedefaultendpunct}{\mcitedefaultseppunct}\relax
\EndOfBibitem
\bibitem[Heinen \emph{et~al.}(2021)Heinen, von Rudorff, and von Lilienfeld]{Heinen2021}
S.~Heinen, G.~F. von Rudorff and O.~A. von Lilienfeld, \emph{J. Chem. Phys.}, 2021, \textbf{155}, 064105\relax
\mciteBstWouldAddEndPuncttrue
\mciteSetBstMidEndSepPunct{\mcitedefaultmidpunct}
{\mcitedefaultendpunct}{\mcitedefaultseppunct}\relax
\EndOfBibitem
\bibitem[Spiekermann \emph{et~al.}(2022)Spiekermann, Pattanaik, and Green]{Spiekermann2022}
K.~A. Spiekermann, L.~Pattanaik and W.~H. Green, \emph{J. Phys. Chem.}, 2022, \textbf{126}, 3976--3986\relax
\mciteBstWouldAddEndPuncttrue
\mciteSetBstMidEndSepPunct{\mcitedefaultmidpunct}
{\mcitedefaultendpunct}{\mcitedefaultseppunct}\relax
\EndOfBibitem
\bibitem[van Gerwen \emph{et~al.}(2024)van Gerwen, Briling, Bunne, Somnath, Laplaza, Krause, and Corminboeuf]{van20243dreact}
P.~van Gerwen, K.~R. Briling, C.~Bunne, V.~R. Somnath, R.~Laplaza, A.~Krause and C.~Corminboeuf, \emph{J. Chem. Inf. Model.}, 2024, \textbf{64}, 5771--5785\relax
\mciteBstWouldAddEndPuncttrue
\mciteSetBstMidEndSepPunct{\mcitedefaultmidpunct}
{\mcitedefaultendpunct}{\mcitedefaultseppunct}\relax
\EndOfBibitem
\bibitem[Duan \emph{et~al.}(2025)Duan, Liu, Du, Chen, Zhao, Jia, Gomes, Theodorou, and Kulik]{Duan2025}
C.~Duan, G.~H. Liu, Y.~Du, T.~Chen, Q.~Zhao, H.~Jia, C.~P. Gomes, E.~A. Theodorou and H.~J. Kulik, \emph{Nat. Mach. Intell.}, 2025, \textbf{7}, 615--626\relax
\mciteBstWouldAddEndPuncttrue
\mciteSetBstMidEndSepPunct{\mcitedefaultmidpunct}
{\mcitedefaultendpunct}{\mcitedefaultseppunct}\relax
\EndOfBibitem
\bibitem[Chang \emph{et~al.}(2025)Chang, Tsai, and Li]{Chang2025}
H.~C. Chang, M.~H. Tsai and Y.~P. Li, \emph{J. Chem. Inf. Model.}, 2025, \textbf{65}, 1367--1377\relax
\mciteBstWouldAddEndPuncttrue
\mciteSetBstMidEndSepPunct{\mcitedefaultmidpunct}
{\mcitedefaultendpunct}{\mcitedefaultseppunct}\relax
\EndOfBibitem
\bibitem[Peters \emph{et~al.}(2004)Peters, Heyden, Bell, and Chakraborty]{Peters2004}
B.~Peters, A.~Heyden, A.~T. Bell and A.~Chakraborty, \emph{J. Chem. Phys}, 2004, \textbf{120}, 7877--7886\relax
\mciteBstWouldAddEndPuncttrue
\mciteSetBstMidEndSepPunct{\mcitedefaultmidpunct}
{\mcitedefaultendpunct}{\mcitedefaultseppunct}\relax
\EndOfBibitem
\bibitem[Behn \emph{et~al.}(2011)Behn, Zimmerman, Bell, and Head-Gordon]{behn2011efficient}
A.~Behn, P.~M. Zimmerman, A.~T. Bell and M.~Head-Gordon, \emph{J. Chem. Phys.}, 2011, \textbf{135}, 224108\relax
\mciteBstWouldAddEndPuncttrue
\mciteSetBstMidEndSepPunct{\mcitedefaultmidpunct}
{\mcitedefaultendpunct}{\mcitedefaultseppunct}\relax
\EndOfBibitem
\bibitem[Sharada \emph{et~al.}(2012)Sharada, Zimmerman, Bell, and Head-Gordon]{MallikarjunSharada2012}
S.~M. Sharada, P.~M. Zimmerman, A.~T. Bell and M.~Head-Gordon, \emph{J. Chem. Theory Comput.}, 2012, \textbf{8}, 5166--5174\relax
\mciteBstWouldAddEndPuncttrue
\mciteSetBstMidEndSepPunct{\mcitedefaultmidpunct}
{\mcitedefaultendpunct}{\mcitedefaultseppunct}\relax
\EndOfBibitem
\bibitem[Zimmerman(2013)]{Zimmerman2013}
P.~M. Zimmerman, \emph{J. Chem. Phys.}, 2013, \textbf{138}, 184102\relax
\mciteBstWouldAddEndPuncttrue
\mciteSetBstMidEndSepPunct{\mcitedefaultmidpunct}
{\mcitedefaultendpunct}{\mcitedefaultseppunct}\relax
\EndOfBibitem
\bibitem[Jafari and Zimmerman(2017)]{jafari2017reliable}
M.~Jafari and P.~M. Zimmerman, \emph{J. Comput. Chem.}, 2017, \textbf{38}, 645--658\relax
\mciteBstWouldAddEndPuncttrue
\mciteSetBstMidEndSepPunct{\mcitedefaultmidpunct}
{\mcitedefaultendpunct}{\mcitedefaultseppunct}\relax
\EndOfBibitem
\bibitem[Mills and J{\'o}nsson(1994)]{mills1994quantum}
G.~Mills and H.~J{\'o}nsson, \emph{Phys. Rev. Lett.}, 1994, \textbf{72}, 1124\relax
\mciteBstWouldAddEndPuncttrue
\mciteSetBstMidEndSepPunct{\mcitedefaultmidpunct}
{\mcitedefaultendpunct}{\mcitedefaultseppunct}\relax
\EndOfBibitem
\bibitem[Henkelman and J{\'o}nsson(2000)]{henkelman2000improved}
G.~Henkelman and H.~J{\'o}nsson, \emph{J. Chem. Phys.}, 2000, \textbf{113}, 9978--9985\relax
\mciteBstWouldAddEndPuncttrue
\mciteSetBstMidEndSepPunct{\mcitedefaultmidpunct}
{\mcitedefaultendpunct}{\mcitedefaultseppunct}\relax
\EndOfBibitem
\bibitem[Henkelman \emph{et~al.}(2000)Henkelman, Uberuaga, and Jonsson]{Henkelman2000climbing}
G.~Henkelman, B.~P. Uberuaga and H.~Jonsson, \emph{J. Chem. Phys.}, 2000, \textbf{113}, 9901--9904\relax
\mciteBstWouldAddEndPuncttrue
\mciteSetBstMidEndSepPunct{\mcitedefaultmidpunct}
{\mcitedefaultendpunct}{\mcitedefaultseppunct}\relax
\EndOfBibitem
\bibitem[Sheppard \emph{et~al.}(2008)Sheppard, Terrell, and Henkelman]{sheppard2008optimization}
D.~Sheppard, R.~Terrell and G.~Henkelman, \emph{J. Chem. Phys}, 2008, \textbf{128}, 134106\relax
\mciteBstWouldAddEndPuncttrue
\mciteSetBstMidEndSepPunct{\mcitedefaultmidpunct}
{\mcitedefaultendpunct}{\mcitedefaultseppunct}\relax
\EndOfBibitem
\bibitem[Marks and Gomes(2025)]{Marks2024incorporation}
J.~Marks and J.~Gomes, \emph{J. Chem. Theory Comput.}, 2025, \textbf{21}, 12110--12120\relax
\mciteBstWouldAddEndPuncttrue
\mciteSetBstMidEndSepPunct{\mcitedefaultmidpunct}
{\mcitedefaultendpunct}{\mcitedefaultseppunct}\relax
\EndOfBibitem
\bibitem[Anstine \emph{et~al.}(2025)Anstine, Zubatyuk, and Isayev]{Anstine2025}
D.~M. Anstine, R.~Zubatyuk and O.~Isayev, \emph{Chem. Sci.}, 2025, \textbf{16}, 10228--10244\relax
\mciteBstWouldAddEndPuncttrue
\mciteSetBstMidEndSepPunct{\mcitedefaultmidpunct}
{\mcitedefaultendpunct}{\mcitedefaultseppunct}\relax
\EndOfBibitem
\bibitem[Batatia \emph{et~al.}(2022)Batatia, Kov{\'a}cs, Simm, Ortner, and Cs{\'a}nyi]{batatia2022mace}
I.~Batatia, D.~P. Kov{\'a}cs, G.~Simm, C.~Ortner and G.~Cs{\'a}nyi, \emph{Adv. Neural Inf. Process. Syst.}, 2022, \textbf{35}, 11423--11436\relax
\mciteBstWouldAddEndPuncttrue
\mciteSetBstMidEndSepPunct{\mcitedefaultmidpunct}
{\mcitedefaultendpunct}{\mcitedefaultseppunct}\relax
\EndOfBibitem
\bibitem[Fu \emph{et~al.}(2025)Fu, Wood, Barroso-Luque, Levine, Gao, Dzamba, and Zitnick]{Fu2025}
X.~Fu, B.~M. Wood, L.~Barroso-Luque, D.~S. Levine, M.~Gao, M.~Dzamba and C.~L. Zitnick, \emph{Proc. Mach. Learn. Res.}, 2025, \textbf{267}, 17875--17893\relax
\mciteBstWouldAddEndPuncttrue
\mciteSetBstMidEndSepPunct{\mcitedefaultmidpunct}
{\mcitedefaultendpunct}{\mcitedefaultseppunct}\relax
\EndOfBibitem
\bibitem[Wood \emph{et~al.}(2025)Wood, Dzamba, Fu, Gao, Shuaibi, Barroso-Luque, Abdelmaqsoud, Gharakhanyan, Kitchin, Levine, Michel, Sriram, Cohen, Das, Sahoo, Rizvi, Ulissi, and Zitnick]{wood2025family}
B.~M. Wood, M.~Dzamba, X.~Fu, M.~Gao, M.~Shuaibi, L.~Barroso-Luque, K.~Abdelmaqsoud, V.~Gharakhanyan, J.~R. Kitchin, D.~S. Levine, K.~Michel, A.~Sriram, T.~Cohen, A.~Das, S.~J. Sahoo, A.~Rizvi, Z.~W. Ulissi and C.~L. Zitnick, \emph{Adv. Neural Inf. Process. Syst.}, 2025\relax
\mciteBstWouldAddEndPuncttrue
\mciteSetBstMidEndSepPunct{\mcitedefaultmidpunct}
{\mcitedefaultendpunct}{\mcitedefaultseppunct}\relax
\EndOfBibitem
\bibitem[Levine \emph{et~al.}(2025)Levine, Shuaibi, Spotte-Smith, Taylor, Hasyim, Michel, Batatia, Csányi, Dzamba, Eastman, Frey, Fu, Gharakhanyan, Krishnapriyan, Rackers, Raja, Rizvi, Rosen, Ulissi, Vargas, Zitnick, Blau, and Wood]{Levine2025}
D.~S. Levine, M.~Shuaibi, E.~W.~C. Spotte-Smith, M.~G. Taylor, M.~R. Hasyim, K.~Michel, I.~Batatia, G.~Csányi, M.~Dzamba, P.~Eastman, N.~C. Frey, X.~Fu, V.~Gharakhanyan, A.~S. Krishnapriyan, J.~A. Rackers, S.~Raja, A.~Rizvi, A.~S. Rosen, Z.~Ulissi, S.~Vargas, C.~L. Zitnick, S.~M. Blau and B.~M. Wood, \emph{arXiv preprint arXiv:2505.08762}, 2025\relax
\mciteBstWouldAddEndPuncttrue
\mciteSetBstMidEndSepPunct{\mcitedefaultmidpunct}
{\mcitedefaultendpunct}{\mcitedefaultseppunct}\relax
\EndOfBibitem
\bibitem[Schreiner \emph{et~al.}(2022)Schreiner, Bhowmik, Vegge, Jørgensen, and Winther]{Schreiner2022}
M.~Schreiner, A.~Bhowmik, T.~Vegge, P.~B. Jørgensen and O.~Winther, \emph{Mach. Learn.: Sci. Technol.}, 2022, \textbf{3}, 045022\relax
\mciteBstWouldAddEndPuncttrue
\mciteSetBstMidEndSepPunct{\mcitedefaultmidpunct}
{\mcitedefaultendpunct}{\mcitedefaultseppunct}\relax
\EndOfBibitem
\bibitem[Wander \emph{et~al.}(2025)Wander, Shuaibi, Kitchin, Ulissi, and Zitnick]{Wander2024}
B.~Wander, M.~Shuaibi, J.~R. Kitchin, Z.~W. Ulissi and C.~L. Zitnick, \emph{ACS Catal.}, 2025, \textbf{15}, 5283--5294\relax
\mciteBstWouldAddEndPuncttrue
\mciteSetBstMidEndSepPunct{\mcitedefaultmidpunct}
{\mcitedefaultendpunct}{\mcitedefaultseppunct}\relax
\EndOfBibitem
\bibitem[Zhao \emph{et~al.}(2025)Zhao, Han, Zhang, Wang, Zhong, Cui, Yin, Cao, Jia, and Duan]{Zhao2025}
Q.~Zhao, Y.~Han, D.~Zhang, J.~Wang, P.~Zhong, T.~Cui, B.~Yin, Y.~Cao, H.~Jia and C.~Duan, \emph{Adv. Sci.}, 2025\relax
\mciteBstWouldAddEndPuncttrue
\mciteSetBstMidEndSepPunct{\mcitedefaultmidpunct}
{\mcitedefaultendpunct}{\mcitedefaultseppunct}\relax
\EndOfBibitem
\bibitem[Marks and Gomes(2025)]{Marks2025}
J.~Marks and J.~Gomes, \emph{arXiv preprint arXiv:2501.06159}, 2025\relax
\mciteBstWouldAddEndPuncttrue
\mciteSetBstMidEndSepPunct{\mcitedefaultmidpunct}
{\mcitedefaultendpunct}{\mcitedefaultseppunct}\relax
\EndOfBibitem
\bibitem[Iron and Janes(2019)]{Iron2019}
M.~A. Iron and T.~Janes, \emph{J. Phys. Chem. A}, 2019, \textbf{123}, 3761--3781\relax
\mciteBstWouldAddEndPuncttrue
\mciteSetBstMidEndSepPunct{\mcitedefaultmidpunct}
{\mcitedefaultendpunct}{\mcitedefaultseppunct}\relax
\EndOfBibitem
\bibitem[Bursch \emph{et~al.}(2022)Bursch, Mewes, Hansen, and Grimme]{bursch2022best}
M.~Bursch, J.-M. Mewes, A.~Hansen and S.~Grimme, \emph{Angew. Chem.}, 2022, \textbf{134}, e202205735\relax
\mciteBstWouldAddEndPuncttrue
\mciteSetBstMidEndSepPunct{\mcitedefaultmidpunct}
{\mcitedefaultendpunct}{\mcitedefaultseppunct}\relax
\EndOfBibitem
\bibitem[Larsen \emph{et~al.}(2017)Larsen, Mortensen, Blomqvist, Castelli, Christensen, Du{\l}ak, Friis, Groves, Hammer, Hargus,\emph{et~al.}]{larsen2017atomic}
A.~H. Larsen, J.~J. Mortensen, J.~Blomqvist, I.~E. Castelli, R.~Christensen, M.~Du{\l}ak, J.~Friis, M.~N. Groves, B.~Hammer, C.~Hargus \emph{et~al.}, \emph{J. Phys.:Condens. Matter}, 2017, \textbf{29}, 273002\relax
\mciteBstWouldAddEndPuncttrue
\mciteSetBstMidEndSepPunct{\mcitedefaultmidpunct}
{\mcitedefaultendpunct}{\mcitedefaultseppunct}\relax
\EndOfBibitem
\bibitem[Hermes \emph{et~al.}(2019)Hermes, Sargsyan, Najm, and Zádor]{Hermes2019}
E.~D. Hermes, K.~Sargsyan, H.~N. Najm and J.~Zádor, \emph{J. Chem. Theory Comput.}, 2019, \textbf{15}, 6536--6549\relax
\mciteBstWouldAddEndPuncttrue
\mciteSetBstMidEndSepPunct{\mcitedefaultmidpunct}
{\mcitedefaultendpunct}{\mcitedefaultseppunct}\relax
\EndOfBibitem
\bibitem[Hermes \emph{et~al.}(2022)Hermes, Sargsyan, Najm, and Zádor]{Hermes2022}
E.~D. Hermes, K.~Sargsyan, H.~N. Najm and J.~Zádor, \emph{J. Chem. Theory Comput.}, 2022, \textbf{18}, 6974--6988\relax
\mciteBstWouldAddEndPuncttrue
\mciteSetBstMidEndSepPunct{\mcitedefaultmidpunct}
{\mcitedefaultendpunct}{\mcitedefaultseppunct}\relax
\EndOfBibitem
\bibitem[Liu and Nocedal(1989)]{liu1989limited}
D.~C. Liu and J.~Nocedal, \emph{Math. Program.}, 1989, \textbf{45}, 503--528\relax
\mciteBstWouldAddEndPuncttrue
\mciteSetBstMidEndSepPunct{\mcitedefaultmidpunct}
{\mcitedefaultendpunct}{\mcitedefaultseppunct}\relax
\EndOfBibitem
\bibitem[Byrd \emph{et~al.}(1995)Byrd, Lu, Nocedal, and Zhu]{byrd1995limited}
R.~H. Byrd, P.~Lu, J.~Nocedal and C.~Zhu, \emph{SIAM J. Sci. Comput.}, 1995, \textbf{16}, 1190--1208\relax
\mciteBstWouldAddEndPuncttrue
\mciteSetBstMidEndSepPunct{\mcitedefaultmidpunct}
{\mcitedefaultendpunct}{\mcitedefaultseppunct}\relax
\EndOfBibitem
\bibitem[Virtanen \emph{et~al.}(2020)Virtanen, Gommers, Oliphant, Haberland, Reddy, Cournapeau, Burovski, Peterson, Weckesser, Bright,\emph{et~al.}]{virtanen2020scipy}
P.~Virtanen, R.~Gommers, T.~E. Oliphant, M.~Haberland, T.~Reddy, D.~Cournapeau, E.~Burovski, P.~Peterson, W.~Weckesser, J.~Bright \emph{et~al.}, \emph{Nat. Methods}, 2020, \textbf{17}, 261--272\relax
\mciteBstWouldAddEndPuncttrue
\mciteSetBstMidEndSepPunct{\mcitedefaultmidpunct}
{\mcitedefaultendpunct}{\mcitedefaultseppunct}\relax
\EndOfBibitem
\bibitem[Bannwarth \emph{et~al.}(2019)Bannwarth, Ehlert, and Grimme]{bannwarth2019gfn2}
C.~Bannwarth, S.~Ehlert and S.~Grimme, \emph{J. Chem. Theory Comput.}, 2019, \textbf{15}, 1652--1671\relax
\mciteBstWouldAddEndPuncttrue
\mciteSetBstMidEndSepPunct{\mcitedefaultmidpunct}
{\mcitedefaultendpunct}{\mcitedefaultseppunct}\relax
\EndOfBibitem
\bibitem[Chanussot \emph{et~al.}(2021)Chanussot, Das, Goyal, Lavril, Shuaibi, Riviere, Tran, Heras-Domingo, Ho, Hu,\emph{et~al.}]{Chanussot2021}
L.~Chanussot, A.~Das, S.~Goyal, T.~Lavril, M.~Shuaibi, M.~Riviere, K.~Tran, J.~Heras-Domingo, C.~Ho, W.~Hu \emph{et~al.}, \emph{ACS Catal.}, 2021, \textbf{11}, 6059--6072\relax
\mciteBstWouldAddEndPuncttrue
\mciteSetBstMidEndSepPunct{\mcitedefaultmidpunct}
{\mcitedefaultendpunct}{\mcitedefaultseppunct}\relax
\EndOfBibitem
\bibitem[Barroso-Luque \emph{et~al.}(2024)Barroso-Luque, Shuaibi, Fu, Wood, Dzamba, Gao, Rizvi, Zitnick, and Ulissi]{Barroso-Luque2024}
L.~Barroso-Luque, M.~Shuaibi, X.~Fu, B.~M. Wood, M.~Dzamba, M.~Gao, A.~Rizvi, C.~L. Zitnick and Z.~W. Ulissi, \emph{arXiv preprint arXiv:2410.12771}, 2024\relax
\mciteBstWouldAddEndPuncttrue
\mciteSetBstMidEndSepPunct{\mcitedefaultmidpunct}
{\mcitedefaultendpunct}{\mcitedefaultseppunct}\relax
\EndOfBibitem
\bibitem[Sriram \emph{et~al.}(2024)Sriram, Choi, Yu, Brabson, Das, Ulissi, Uyttendaele, Medford, and Sholl]{Sriram2024}
A.~Sriram, S.~Choi, X.~Yu, L.~M. Brabson, A.~Das, Z.~Ulissi, M.~Uyttendaele, A.~J. Medford and D.~S. Sholl, \emph{ACS Cent. Sci.}, 2024, \textbf{10}, 923--941\relax
\mciteBstWouldAddEndPuncttrue
\mciteSetBstMidEndSepPunct{\mcitedefaultmidpunct}
{\mcitedefaultendpunct}{\mcitedefaultseppunct}\relax
\EndOfBibitem
\bibitem[Epifanovsky \emph{et~al.}(2021)Epifanovsky, Gilbert, Feng, Lee, Mao, Mardirossian, Pokhilko, White, Coons, Dempwolff,\emph{et~al.}]{Epifanovsky2021}
E.~Epifanovsky, A.~T. Gilbert, X.~Feng, J.~Lee, Y.~Mao, N.~Mardirossian, P.~Pokhilko, A.~F. White, M.~P. Coons, A.~L. Dempwolff \emph{et~al.}, \emph{J. Chem. Phys.}, 2021, \textbf{155}, 084801\relax
\mciteBstWouldAddEndPuncttrue
\mciteSetBstMidEndSepPunct{\mcitedefaultmidpunct}
{\mcitedefaultendpunct}{\mcitedefaultseppunct}\relax
\EndOfBibitem
\bibitem[Mardirossian and Head-Gordon(2014)]{Mardirossian2014}
N.~Mardirossian and M.~Head-Gordon, \emph{Phys. Chem. Chem. Phys.}, 2014, \textbf{16}, 9904--9924\relax
\mciteBstWouldAddEndPuncttrue
\mciteSetBstMidEndSepPunct{\mcitedefaultmidpunct}
{\mcitedefaultendpunct}{\mcitedefaultseppunct}\relax
\EndOfBibitem
\bibitem[Weigend and Ahlrichs(2005)]{Weigend2005}
F.~Weigend and R.~Ahlrichs, \emph{Phys. Chem. Chem. Phys.}, 2005,  3297--3305\relax
\mciteBstWouldAddEndPuncttrue
\mciteSetBstMidEndSepPunct{\mcitedefaultmidpunct}
{\mcitedefaultendpunct}{\mcitedefaultseppunct}\relax
\EndOfBibitem
\bibitem[Dasgupta and Herbert(2017)]{dasgupta2017standard}
S.~Dasgupta and J.~M. Herbert, \emph{J. Comput. Chem.}, 2017, \textbf{38}, 869--882\relax
\mciteBstWouldAddEndPuncttrue
\mciteSetBstMidEndSepPunct{\mcitedefaultmidpunct}
{\mcitedefaultendpunct}{\mcitedefaultseppunct}\relax
\EndOfBibitem
\bibitem[Baker(1986)]{Baker1986}
J.~Baker, \emph{J. Comput. Chem.}, 1986, \textbf{7}, 385--395\relax
\mciteBstWouldAddEndPuncttrue
\mciteSetBstMidEndSepPunct{\mcitedefaultmidpunct}
{\mcitedefaultendpunct}{\mcitedefaultseppunct}\relax
\EndOfBibitem
\bibitem[Schmidt \emph{et~al.}(1985)Schmidt, Gordon, and Dupuis]{schmidt1985intrinsic}
M.~W. Schmidt, M.~S. Gordon and M.~Dupuis, \emph{J. Am. Chem. Soc.}, 1985, \textbf{107}, 2585--2589\relax
\mciteBstWouldAddEndPuncttrue
\mciteSetBstMidEndSepPunct{\mcitedefaultmidpunct}
{\mcitedefaultendpunct}{\mcitedefaultseppunct}\relax
\EndOfBibitem
\bibitem[Hermes \emph{et~al.}(2021)Hermes, Sargsyan, Najm, and Z{\'a}dor]{hermes2021geometry}
E.~D. Hermes, K.~Sargsyan, H.~N. Najm and J.~Z{\'a}dor, \emph{J. Chem. Phys.}, 2021, \textbf{155}, 094105\relax
\mciteBstWouldAddEndPuncttrue
\mciteSetBstMidEndSepPunct{\mcitedefaultmidpunct}
{\mcitedefaultendpunct}{\mcitedefaultseppunct}\relax
\EndOfBibitem
\bibitem[Murtagh and Sargent(1970)]{murtagh1970computational}
B.~A. Murtagh and R.~W. Sargent, \emph{Comput. J.}, 1970, \textbf{13}, 185--194\relax
\mciteBstWouldAddEndPuncttrue
\mciteSetBstMidEndSepPunct{\mcitedefaultmidpunct}
{\mcitedefaultendpunct}{\mcitedefaultseppunct}\relax
\EndOfBibitem
\bibitem[Baker and Chan(1996)]{baker1996location}
J.~Baker and F.~Chan, \emph{J. Comput. Chem.}, 1996, \textbf{17}, 888--904\relax
\mciteBstWouldAddEndPuncttrue
\mciteSetBstMidEndSepPunct{\mcitedefaultmidpunct}
{\mcitedefaultendpunct}{\mcitedefaultseppunct}\relax
\EndOfBibitem
\bibitem[Heyden \emph{et~al.}(2005)Heyden, Bell, and Keil]{Heyden2005}
A.~Heyden, A.~T. Bell and F.~J. Keil, \emph{J. Chem. Phys.}, 2005, \textbf{123}, 224101\relax
\mciteBstWouldAddEndPuncttrue
\mciteSetBstMidEndSepPunct{\mcitedefaultmidpunct}
{\mcitedefaultendpunct}{\mcitedefaultseppunct}\relax
\EndOfBibitem
\bibitem[K{\"a}stner and Sherwood(2008)]{kastner2008superlinearly}
J.~K{\"a}stner and P.~Sherwood, \emph{J. Chem. Phys.}, 2008, \textbf{128}, 014106\relax
\mciteBstWouldAddEndPuncttrue
\mciteSetBstMidEndSepPunct{\mcitedefaultmidpunct}
{\mcitedefaultendpunct}{\mcitedefaultseppunct}\relax
\EndOfBibitem
\bibitem[Hait \emph{et~al.}(2025)Hait, Estrada~Pabon, Stohr, and Mart{\'\i}nez]{Hait2025}
D.~Hait, J.~D. Estrada~Pabon, M.~Stohr and T.~J. Mart{\'\i}nez, \emph{J. Chem. Theory Comput.}, 2025, \textbf{21}, 11632--11644\relax
\mciteBstWouldAddEndPuncttrue
\mciteSetBstMidEndSepPunct{\mcitedefaultmidpunct}
{\mcitedefaultendpunct}{\mcitedefaultseppunct}\relax
\EndOfBibitem
\bibitem[Halgren and Lipscomb(1977)]{Halgren1977}
T.~A. Halgren and W.~N. Lipscomb, \emph{Chem. Phys. Lett.}, 1977, \textbf{49}, 225--232\relax
\mciteBstWouldAddEndPuncttrue
\mciteSetBstMidEndSepPunct{\mcitedefaultmidpunct}
{\mcitedefaultendpunct}{\mcitedefaultseppunct}\relax
\EndOfBibitem
\bibitem[Maeda \emph{et~al.}(2016)Maeda, Harabuchi, Takagi, Taketsugu, and Morokuma]{maeda2016artificial}
S.~Maeda, Y.~Harabuchi, M.~Takagi, T.~Taketsugu and K.~Morokuma, \emph{Chem. Rec.}, 2016, \textbf{16}, 2232--2248\relax
\mciteBstWouldAddEndPuncttrue
\mciteSetBstMidEndSepPunct{\mcitedefaultmidpunct}
{\mcitedefaultendpunct}{\mcitedefaultseppunct}\relax
\EndOfBibitem
\bibitem[Zakzeski \emph{et~al.}(2009)Zakzeski, Behn, Head-Gordon, and Bell]{Zakzeski2009}
J.~Zakzeski, A.~Behn, M.~Head-Gordon and A.~T. Bell, \emph{J. Am. Chem. Soc.}, 2009, \textbf{131}, 11098--11105\relax
\mciteBstWouldAddEndPuncttrue
\mciteSetBstMidEndSepPunct{\mcitedefaultmidpunct}
{\mcitedefaultendpunct}{\mcitedefaultseppunct}\relax
\EndOfBibitem
\bibitem[Liu \emph{et~al.}(2017)Liu, Luo, Hou, and Luo]{Liu2017}
F.~Liu, G.~Luo, Z.~Hou and Y.~Luo, \emph{Organometallics}, 2017, \textbf{36}, 1557--1565\relax
\mciteBstWouldAddEndPuncttrue
\mciteSetBstMidEndSepPunct{\mcitedefaultmidpunct}
{\mcitedefaultendpunct}{\mcitedefaultseppunct}\relax
\EndOfBibitem
\bibitem[Prokopchuk and Morris(2012)]{Prokopchuk2012}
D.~E. Prokopchuk and R.~H. Morris, \emph{Organometallics}, 2012, \textbf{31}, 7375--7385\relax
\mciteBstWouldAddEndPuncttrue
\mciteSetBstMidEndSepPunct{\mcitedefaultmidpunct}
{\mcitedefaultendpunct}{\mcitedefaultseppunct}\relax
\EndOfBibitem
\bibitem[Iron \emph{et~al.}(2003)Iron, Martin, and der Boom]{Iron2003}
M.~A. Iron, J.~M. Martin and M.~E.~V. der Boom, \emph{J. Am. Chem. Soc.}, 2003, \textbf{125}, 13020--13021\relax
\mciteBstWouldAddEndPuncttrue
\mciteSetBstMidEndSepPunct{\mcitedefaultmidpunct}
{\mcitedefaultendpunct}{\mcitedefaultseppunct}\relax
\EndOfBibitem
\bibitem[Sriram \emph{et~al.}(2022)Sriram, Das, Wood, Goyal, and Zitnick]{Sriram2022}
A.~Sriram, A.~Das, B.~M. Wood, S.~Goyal and C.~L. Zitnick, \emph{Proc. Int. Conf. Learn. Represent.}, 2022\relax
\mciteBstWouldAddEndPuncttrue
\mciteSetBstMidEndSepPunct{\mcitedefaultmidpunct}
{\mcitedefaultendpunct}{\mcitedefaultseppunct}\relax
\EndOfBibitem
\bibitem[Tran \emph{et~al.}(2023)Tran, Lan, Shuaibi, Wood, Goyal, Das, Heras-Domingo, Kolluru, Rizvi, Shoghi,\emph{et~al.}]{Tran2023}
R.~Tran, J.~Lan, M.~Shuaibi, B.~M. Wood, S.~Goyal, A.~Das, J.~Heras-Domingo, A.~Kolluru, A.~Rizvi, N.~Shoghi \emph{et~al.}, \emph{ACS Catal.}, 2023, \textbf{13}, 3066--3084\relax
\mciteBstWouldAddEndPuncttrue
\mciteSetBstMidEndSepPunct{\mcitedefaultmidpunct}
{\mcitedefaultendpunct}{\mcitedefaultseppunct}\relax
\EndOfBibitem
\bibitem[Powell(1971)]{powell1971recent}
M.~J. Powell, \emph{Math. Program.}, 1971, \textbf{1}, 26--57\relax
\mciteBstWouldAddEndPuncttrue
\mciteSetBstMidEndSepPunct{\mcitedefaultmidpunct}
{\mcitedefaultendpunct}{\mcitedefaultseppunct}\relax
\EndOfBibitem
\bibitem[Yuan \emph{et~al.}(2024)Yuan, Kumar, Guan, Hermes, Rosen, Z{\'a}dor, Head-Gordon, and Blau]{yuan2024analytical}
E.~C.-Y. Yuan, A.~Kumar, X.~Guan, E.~D. Hermes, A.~S. Rosen, J.~Z{\'a}dor, T.~Head-Gordon and S.~M. Blau, \emph{Nat. Commun.}, 2024, \textbf{15}, 8865\relax
\mciteBstWouldAddEndPuncttrue
\mciteSetBstMidEndSepPunct{\mcitedefaultmidpunct}
{\mcitedefaultendpunct}{\mcitedefaultseppunct}\relax
\EndOfBibitem
\bibitem[Burger \emph{et~al.}(2025)Burger, Thiede, R{\o}nne, Bernales, Vijaykumar, Vegge, Bhowmik, and Aspuru-Guzik]{burger2025shoot}
A.~Burger, L.~Thiede, N.~R{\o}nne, V.~Bernales, N.~Vijaykumar, T.~Vegge, A.~Bhowmik and A.~Aspuru-Guzik, \emph{arXiv preprint arXiv:2509.21624}, 2025\relax
\mciteBstWouldAddEndPuncttrue
\mciteSetBstMidEndSepPunct{\mcitedefaultmidpunct}
{\mcitedefaultendpunct}{\mcitedefaultseppunct}\relax
\EndOfBibitem
\bibitem[Bronkema \emph{et~al.}(2007)Bronkema, Leo, and Bell]{Bronkema2007}
J.~L. Bronkema, D.~C. Leo and A.~T. Bell, \emph{J. Phys. Chem. C}, 2007, \textbf{111}, 14530--14540\relax
\mciteBstWouldAddEndPuncttrue
\mciteSetBstMidEndSepPunct{\mcitedefaultmidpunct}
{\mcitedefaultendpunct}{\mcitedefaultseppunct}\relax
\EndOfBibitem
\bibitem[Goodrow and Bell(2008)]{Goodrow2008}
A.~Goodrow and A.~T. Bell, \emph{J. Phys. Chem. C}, 2008, \textbf{112}, 13204--13214\relax
\mciteBstWouldAddEndPuncttrue
\mciteSetBstMidEndSepPunct{\mcitedefaultmidpunct}
{\mcitedefaultendpunct}{\mcitedefaultseppunct}\relax
\EndOfBibitem
\bibitem[Makri \emph{et~al.}(2019)Makri, Ortner, and Kermode]{makri2019preconditioning}
S.~Makri, C.~Ortner and J.~R. Kermode, \emph{J. Chem. Phys.}, 2019, \textbf{150}, 094109\relax
\mciteBstWouldAddEndPuncttrue
\mciteSetBstMidEndSepPunct{\mcitedefaultmidpunct}
{\mcitedefaultendpunct}{\mcitedefaultseppunct}\relax
\EndOfBibitem
\end{mcitethebibliography}
\bibliographystyle{rsc} 

\appendix
\section{Appendix A: Nudged Elastic Band Experiments}\label{sec:appendix}
\subsection{Methods}
\label{sec:methods_CI-NEB}
The Nudged Elastic Band (NEB) method locates minimum-energy pathways and approximate transition states by representing the reaction coordinate as a series of intermediate geometries or images $\{\textbf{R}_{i}\}$, connecting reactant and product structures.\cite{mills1994quantum,henkelman2000improved,sheppard2008optimization} In contrast to more common IDPP-based interpolation schemes, 
the initial reaction path in this work was constructed using internal coordinates, which better preserve realistic bonding environments for molecular systems.\cite{Marks2024incorporation} Adjacent images are linked by spring constants of magnitude $k$ enforcing approximately equal spacing along the path, while interatomic forces from the potential-energy function optimize each node perpendicular to the path tangent. For each image $i$ the total force

\begin{equation}
\mathbf{F}_i = \mathbf{F}_i^\perp + \mathbf{F}_i^\parallel
\label{eq:neb_fi}
\end{equation}

is decomposed into perpendicular and parallel components:

\begin{equation}
\mathbf{F}_i^\perp = -\nabla V(\mathbf{R}_i) + [\nabla V(\mathbf{R}_i) \cdot \hat{\mathbf{\tau}}_i]\hat{\mathbf{\tau}}_i
\label{eq:neb_fperp}
\end{equation}

\begin{equation}
\mathbf{F}_i^\parallel = k(|\mathbf{R}_{i+1}-\mathbf{R}_i|-|\textbf{R}_i-\mathbf{R}_{i-1}|)\cdot\hat{\mathbf{\tau}}_i
\label{eq:neb_fparallel}
\end{equation}

where $\nabla V$ denotes the gradient of the potential energy and $\hat{\mathbf{\tau}}_i$ is the normalized local tangent at image $i$. 

The climbing-image nudged elastic band (CI-NEB) variant was employed to determine saddle-point geometries, avoiding the computational cost associated with Hessian calculations required for local surface-walking algorithms.\cite{Henkelman2000climbing} In this approach, once the NEB method converges to a specified force threshold, the highest energy image $j$ is identified and optimized along the reaction coordinate to ascend directly to the transition state. The spring force is removed for this image, and its parallel force component is inverted, yielding:

\begin{equation}
\mathbf{F}_j^\text{CI} = -\nabla V(\mathbf{R}_j)+2[\nabla V(\mathbf{R}_j)\cdot \hat{\tau}_j]\hat{\tau}
\label{eq:CI_fj}
\end{equation}

This drives the image uphill along the path while continuing to relax in the perpendicular direction.

To ensure robust convergence CI-NEB calculations were performed in two sequential steps. First, the path was optimized using the standard NEB method until the maximal perpendicular force to the MEP fell below 0.5 eV$\cdot$\AA$^{-1}$. Subsequently, starting from this relaxed path, the CI-NEB method was then applied until full convergence was achieved with maximal perpendicular forces below 0.05 eV \AA$^{-1}$ (approximately $3\times10^{-3}$ Ha$\cdot$Bohr$^{-1}$). All calculations employed seven nodes along each reaction path with spring constants of 0.1 eV$\cdot$\AA$^{-1}$, using the adaptive step-size-preconditioned NEB optimizer\cite{makri2019preconditioning} as implemented in ASE.\cite{larsen2017atomic}

\subsection{Experiments}
The CI-NEB method exhibited substantially lower success rates than the FSM (Section~\ref{sec:fsm_organic}) across both native and low-level-refined workflows.  On the Baker benchmark, the CI-NEB method succeeded on only 70.8\% of reactions across both workflows. The corresponding average gradient evaluations were 5.6 for native searches and 4.3 for low-level-refined searches, indicating that pre-optimization on the low-level surface does not appreciably lower the downstream high-level refinement cost. This is due to the climbing-image formulation itself, which explicitly maximizes the energy of the highest node in the path. On the Sharada benchmark, the CI-NEB method again showed reduced reliability: across nine reactions, it succeeded on 63\% of native searches and 61.1\% of low-level-refined searches. The average cost of the successful runs was 12.6 and 8.0 gradient evaluations, respectively. 

CI-NEB guesses frequently converged to local minima rather than the target transition state, or failed to converge within 250 optimization cycles of the high-level P-RFO. These failure modes are examined in greater detail in \ref{sec:discussion}. Representative examples include reactions 7 (formyloxyethyl migration), 8 (Diels–Alder cycloaddition), and 15 (H$_2$O + PO$_3^-$ $\rightarrow$ H$_2$PO$_4^-$) of the Baker set. Although the overall performance reflects both the algorithmic strategy, underlying PES, and P-RFO performance, the consistently higher success rates achieved by the FSM on these same systems suggest this implementation of CI-NEB is intrinsically less robust for generating transition-state guesses suitable for high-level DFT optimization. Nevertheless, the low computational cost is noteworthy: when the CI-NEB method does succeed the resulting geometry is often nearly identical to the reference transition state, effectively precluding the need for additional low-level refinement. 

\begin{sidewaystable*}
    \centering
    \caption{Comparison of performance of GFN2-xTB, AIMNet2, eSEN-S, UMA-S, UMA-M, MACE-OMol25 for native and low-level-refined guess generation via the CI-NEB method on the Baker set. Performance is measured by successful convergence to the reference transition state, and the number of DFT gradient evaluations required. Italicized values denote failed runs, with superscripts denoting the failure mode.}
    \label{tab:NEB_baker}
    \resizebox{\textwidth}{!}{\begin{tabular}{l cc cc cc cc cc cc}
        \toprule
         & \multicolumn{2}{c}{GFN2-xTB} & \multicolumn{2}{c}{AIMNet2} & \multicolumn{2}{c}{eSEN-S} & \multicolumn{2}{c}{UMA-S} & \multicolumn{2}{c}{UMA-M} & \multicolumn{2}{c}{MACE-OMol25} \\
         &  Native & Low-level &  Native & Low-level &  Native & Low-level &  Native & Low-level &  Native & Low-level &  Native & Low-level \\
         Reaction & &refined& &refined& &refined& &refined& &refined& &refined\\
        \midrule
        1. HCN $\rightarrow$ HNC  & 4 & 4 & 4 & 4 & 2 & 2 & 3 & 3 & 2 & 2 & 2 & 2 \\
        2. HCCH $\rightarrow$ CCH$_2$ & 4 & 5 & 3 & 3 & 2 & 2 & 2 & 2 & 2 & 2 & 2 & 3 \\
        3. H$_2$CO $\rightarrow$ H$_2$ + CO & 5 & 5 & 12 & 14 & 2 & 2 & \textit{11}$^b$ & \textit{12}$^b$ & \textit{38}$^a$ & \textit{37}$^a$ & 2 & 2 \\
        4. CH$_3$O $\rightarrow$ CH$_2$OH & 4 & 4 & 15 & 14 & 3 & 3 & 3 & 3 & 3 & 3 & 3 & 3 \\
        5. cyclopropyl ring opening & 4 & 5 & \textit{11}$^a$ &  \textit{7}$^a$ & 3 & 3 & 4 & 3 & 3 & 3 & 12 & 3 \\
        6. bicyclo[1.1.0]butane $\rightarrow$ $\textit{trans}$-butadiene  & 19 & 14 & \textit{27}$^b$ &  \textit{4}$^a$ & 4 & 4 & 4 & 4 & 4 & 3 & 3 & 4 \\
        7. formyloxyethyl 1,2-migration & 8 & 8 & \textit{24}$^b$ &  \textit{4}$^b$ &  \textit{107}$^b$ &  \textit{154}$^b$ &  \textit{250}$^d$ &  \textit{188}$^b$ & \textit{81}$^e$ &  \textit{241}$^b$ &  \textit{166}$^b$ &  \textit{250}$^d$ \\
        8. parent Diels–Alder cycloaddition & \textit{17}$^{b}$ &  \textit{167}$^b$ &  \textit{250}$^d$ &  \textit{250}$^d$ &  \textit{250}$^d$ &  \textit{250}$^d$ &  \textit{250}$^d$ &  \textit{250}$^d$ & \textit{71}$^b$ & \textit{62}$^b$ & \textit{79}$^b$ &  \textit{5}$^b$ \\
        9. s-tetrazine $\rightarrow$ 2HCN + N$_2$ & 12 & 12 & 4 & 4 & \textit{74}$^f$ & \textit{49}$^a$ &  \textit{101}$^f$ & 4 & \textit{74}$^f$ &  \textit{139}$^a$ &  \textit{108}$^f$ &  \textit{177}$^f$ \\
        10. $\textit{trans}$-butadiene $\rightarrow$ $\textit{cis}$-butadiene & 4 & 4 & 4 & 4 & 4 & 3 & 4 & 2 & 4 & 2 & 4 & 2 \\
        11. CH$_3$CH$_3$ $\rightarrow$ CH$_2$CH$_2$ + H$_2$ &  \textit{213}$^a$ &  \textit{250}$^d$ & 35 &  \textit{175}$^a$ & 3 & 3 &  \textit{154}$^a$ & \textit{39}$^a$ & 3 & 3 &  \textit{250}$^d$ &  \textit{138}$^a$ \\
        12. CH$_3$CH$_2$F $\rightarrow$ CH$_2$CH$_2$ + HF & 9 & 8 & 4 & 4 & 3 & 2 & 3 & 2 & 2 & 2 & 3 & 2 \\
        13. acetaldehyde keto-enol tautomerism & 5 & 5 & 4 & 4 & 3 & 2 & 3 & 2 & 3 & 2 & 3 & 2 \\
        14. HCOCl $\rightarrow$ HCl + CO & 6 & 6 & 4 & 4 & 2 & 2 & 2 & 2 & 2 & 2 & 2 & 2 \\
        15. H$_2$O + PO$_3^-$ $\rightarrow$ H$_2$PO$_4^-$ &  \textit{250}$^d$ &  \textit{250}$^d$ &  \textit{250}$^d$ &  \textit{1}$^g$ &  \textit{250}$^d$ & \textit{37}$^b$ &  \textit{250}$^d$ &  \textit{250}$^d$ &  \textit{250}$^d$ &  \textit{250}$^d$ &  \textit{157}$^b$ & \textit{20}$^b$ \\
        16. CH$_2$CHCH$_2$CH$_2$CHO Claisen rearrangement & 8 & 7 & 4 & 6 & 3 & 3 & 3 & 3 & 3 & 2 & 3 & 3 \\
        17. SiH$_2$ + CH$_3$CH$_3$ $\rightarrow$ SiH$_3$CH$_2$CH$_3$ & 16 & 15 & 10 & 8 & 5 & 4 & 5 & 5 & 8 & 4 & 5 & 4 \\
        18. HNCCS $\rightarrow$ HNC + CS &  \textit{5}$^g$ &  \textit{107}$^b$ & 2 & 2 & 3 & 3 & 3 & 3 & 3 & 3 & 3 & 2 \\
        19. HCONH$_3^+$ $\rightarrow$ NH$_4^+$ + CO &  \textit{3}$^g$ & 9 & \textit{61}$^c$ & \textit{14}$^b$ & 10 & 9 & 10 & 10 & 10 & 9 & 9 &  \textit{1}$^g$ \\
        20. acrolein rotational TS & 4 & 4 & 3 & 3 & 3 & 2 & 3 & 3 & 3 & 2 & 3 & 3 \\
        21. HCONHOH $\rightarrow$ HCOHNHO & 6 & 10 & 7 & 7 & 4 & 3 & 4 & 3 & 4 & 3 & 4 & 3 \\
        22. HNC + H$_2$ $\rightarrow$ H$_2$CNH &  \textit{7}$^a$ &  \textit{7}$^a$ & \textit{76}$^b$ & \textit{57}$^b$ &  \textit{2}$^a$ &  \textit{2}$^a$ &  \textit{3}$^a$ &  \textit{3}$^a$ &  \textit{2}$^a$ &  \textit{2}$^a$ &  \textit{-}$^h$ &  \textit{-}$^h$ \\ 
        23. H$_2$CNH $\rightarrow$ HCNH$_2$ &  \textit{3}$^f$ &  \textit{3}$^f$ &  \textit{5}$^f$ & 13 & 9 & 2 & 3 & 2 & 9 & 9 &  \textit{2}$^f$ &  \textit{2}$^f$ \\
        24. HCNH$_2$ $\rightarrow$ HCN + H$_2$ & 4 & 4 & 9 & 10 & \textit{17}$^a$ &  \textit{3}$^a$ & \textit{17}$^a$ &  \textit{2}$^a$ & \textit{21}$^f$ &  \textit{250}$^d$ & 4 & 4 \\
        \midrule
        Success Rate & 70.8\% & 75.0\% & 70.8\% & 66.7\% & 75.0\% & 75.0\% & 66.7\% & 70.8\% & 70.8\% & 70.8\% & 70.8\% & 66.7\% \\
        Mean Success Cost & 7.2 & 7.2 & 10.9 & 6.5 & 3.8 & 3.0 & 3.7 & 3.3 & 4.0 & 3.3 & 3.9 & 2.8 \\
        \bottomrule
    \end{tabular}
    }
    \raggedright
    (a) Converges to an alternate first-order saddle point. (b) Converges to local-minimum structure. (c) Converges to correct TS with spurious imaginary frequency additional cost to eliminate frequency via tighter P-RFO convergence parameters was added. (d) Fails to converge P-RFO within 250 optimization cycles. (e) SCF Convergence error. (f) Converges to structure with multiple strong ($>200 \text{ cm}^{-1}$) imaginary frequencies. (g) P-RFO optimization step failure. (h) Q-Chem internal coordinates back transformation failure.\\
\end{sidewaystable*}

\begin{sidewaystable}
    \centering
    \caption{Comparison of performance of GFN2-xTB, AIMNet2, eSEN-S, UMA-S, UMA-M, MACE-OMol25 for native and low-level-refined guess generation via the CI-NEB method on the Sharada set. Performance is measured by successful convergence to the reference transition state, and the number of DFT gradient evaluations required. Italicized values denote failed runs, with superscripts denoting the failure mode.}
    \label{tab:NEB_sharada}
    \resizebox{\textwidth}{!}{\begin{tabular}{l cc cc cc cc cc cc}
        \toprule
         & \multicolumn{2}{c}{GFN2-xTB} & \multicolumn{2}{c}{AIMNet2} & \multicolumn{2}{c}{eSEN-S} & \multicolumn{2}{c}{UMA-S} & \multicolumn{2}{c}{UMA-M} & \multicolumn{2}{c}{MACE-OMol25} \\
         & {Native} & {Low-level} & {Native} & {Low-level} & {Native} & {Low-level} & {Native} & {Low-level} & {Native} & {Low-level} & {Native} & {Low-level} \\
         Reaction & & {refined} & & {refined} & & {refined} & & {refined} & & {refined} & & {refined} \\
        \midrule
        1. H$_2$CO $\rightarrow$ H$_2$ + CO & 5 & 5 & 12 & 14 & 2 & 2 & \textit{11}$^b$ & {\textit{12}$^b$} & {\textit{38}$^a$} & {\textit{37}$^a$} & 2 & 2 \\
        2. SiH$_2$ + H$_2$ $\rightarrow$ SiH$_4$ & 22 & 27 & {\textit{26}$^b$} & {\textit{22}$^b$} & 5 & 4 & 5 & 5 & 5 & 4 & 5 & 4 \\
        3. CH$_2$CHOH $\leftrightarrow$ CH$_3$CHO & 5 & 5 & 4 & 4 & 3 & 2 & 3 & 2 & 3 & 2 & 3 & 2 \\
        4. CH$_3$CH$_3$ $\rightarrow$ CH$_2$CH$_2$ + H$_2$ & {\textit{213}$^a$} & {\textit{250}$^b$} & 35 & {\textit{175}$^a$} & 3 & 3 & {\textit{154}$^a$} & {\textit{39}$^a$} & 3 & 3 & \textit{250}$^d$ & \textit{138}$^a$ \\
        5. bicyclo[1.1.0]butane $\rightarrow$ $\textit{trans}$-butadiene  & 19 & 14 & {\textit{27}$^b$} & {\textit{4}$^a$} & 4 & 4 & 4 & 4 & 4 & 3 & 3 & 4 \\
        6. parent Diels–Alder cycloaddition & {\textit{17}$^b$} & {\textit{167}$^b$} &  {\textit{250}$^d$} & {\textit{250}$^d$} & {\textit{250}$^d$} & {\textit{250}$^d$} & {\textit{250}$^d$} & {\textit{250}$^d$} & {\textit{71}$^b$} & {\textit{62}$^b$} & {\textit{79}$^b$} & {\textit{5}$^b$} \\
        7. cis,cis-2,4-hexadiene $\leftrightarrow$ 3,4-dimethylcyclobutene & {\textit{15}$^b$} & {\textit{12}$^b$} & {\textit{250}$^d$} & {\textit{250}$^d$} & 3 & 3 & 3 & 3 & 3 & 3 & 3 & 3 \\
        8. C$_5$ $\leftrightarrow$ C$_{7AX}$ &  30 &  13 &  42 & {\textit{16}$^c$} &  14 & 8 &  21 &  10 & 14 & 10 &  20 &  11 \\
        9. silyl ketene acetal $\rightarrow$ silyl ester Ireland–Claisen rearrangement & {\textit{62}$^b$} & {\textit{51}$^b$} & {\textit{57}$^b$} & {\textit{54}$^b$} & {\textit{53}$^b$} & {\textit{54}$^b$} & {\textit{250}$^d$} & 65 & 115 & {\textit{250}$^d$} & {\textit{58}$^b$} & {\textit{59}$^b$} \\
        \midrule
        Success Rate & 55.6\% & 55.6\% & 44.4\% & 33.3\% & 77.8\% & 77.8\% & 55.6\%  & 66.7\% & 77.8\% & 66.7\%& 66.7\%& 66.7\% \\
        Mean Success Cost & 16.2 & 12.8 & 23.3 & 11.3 & 4.9 & 3.7 & 7.2 & 14.8 & 21.0 & 4.2 & 6.0 & 4.3 \\
    \bottomrule
    \end{tabular}
    }
    \raggedright
    (a) Converges to an alternate first-order saddle point. (b) Converges to local-minimum structure. (c) Converges to correct TS with spurious imaginary frequency additional cost to eliminate frequency via tighter P-RFO convergence parameters was added. (d) Fails to converge P-RFO within 250 optimization cycles. (e) SCF Convergence error. (f) Converges to structure with multiple strong ($>200 \text{ cm}^{-1}$) imaginary frequencies. (g) P-RFO optimization step failure. \\
\end{sidewaystable}

\subsection{Comparison to the ML-FSM}
\label{sec:FSM_NEB_COMPARE}

Across both benchmark sets a clear algorithmic difference emerges: the FSM consistently outperformed the CI-NEB method in success rate, while achieving a comparable or lower cost when using the low-level refinement. The FSM maintained a high success rate across all reactions and potentials, with no reactions failing for all potentials and workflows attempted. In contrast, the CI-NEB method exhibited lower success rates, with more frequent convergence to local minima or off-target saddle points. The CI-NEB searches fail to produce a transition-state guess that converges to the reference saddle point for reactions 8 (Diels–Alder cycloaddition) and 22 (HNC + H$_2$ $\rightarrow$ H$_2$CNH) of the Baker set. These results demonstrate that the FSM provides a more reliable and computationally efficient framework for generating transition-state guesses. Consequently, the FSM is exclusively used in subsequent benchmark sets and case studies.

\end{document}